\documentclass[aps,pre,reprint,superscriptaddress]{revtex4-1}

\usepackage{graphics}
\usepackage[dvips]{graphicx}
\usepackage[position=top]{subfig}
\usepackage{amsmath}
\usepackage{amsfonts}
\usepackage{hyperref}
\usepackage{algpseudocode}

\usepackage{xcolor}
\usepackage{array}

\definecolor{d0}{rgb}{0,1,0} 
\definecolor{d1}{rgb}{0,1,1} 
\definecolor{d2}{rgb}{0,0,1} 
\definecolor{d3}{rgb}{0.5,0,1} 

\newcommand{\dda}{\textcolor{d0}}
\newcommand{\ddb}{\textcolor{d1}}
\newcommand{\ddc}{\textcolor{d2}}
\newcommand{\ddd}{\textcolor{d3}}

\newcommand{\interaction}{{\mathcal{A}}}
\newcommand{\interactionM}{\mathbf{A}}
\newcommand{\interactionMel}{{A}}

\newcommand{\order}{{o}}
\newcommand{\q}{\mathbf{q}}
\newcommand{\T}{\mathbf{T}}
\newcommand{\G}{\mathbf{G}}
\newcommand{\s}{\mathbf{s}}
\newcommand{\R}{\mathbf{R}}

\begin{document}

\title{Inference of time-ordered multibody interactions}
\date{\today}

\author{Unai Alvarez-Rodriguez}
\affiliation{University of Deusto, Bilbao, Spain}
\affiliation{University of Zurich, Z\"{u}rich, Switzerland}

\author{Luka V. Petrovi\'c}
\affiliation{University of Zurich, Z\"{u}rich, Switzerland}

\author{Ingo Scholtes}
\affiliation{Julius-Maximilians-Universit\"at W\"urzburg, W\"urzburg, Germany}
\affiliation{University of Zurich, Z\"{u}rich, Switzerland}

\begin{abstract}
We introduce time-ordered multibody interactions to describe complex systems manifesting temporal as well as multibody dependencies.
First, we show how the dynamics of multivariate Markov chains can be decomposed in ensembles of time-ordered multibody interactions.
Then, we present an algorithm to extract  those interactions from data capturing the system-level dynamics of node states and a measure to characterize the complexity of interaction ensembles. 
Finally, we experimentally validate the robustness of our algorithm against statistical errors and its efficiency at inferring parsimonious interaction ensembles.
\end{abstract}

\maketitle

\section{Introduction}

Earth's climate, traffic in large cities, or a baroque musical composition are all examples of complex systems.
They are composed of multiple elements with different types of interactions and elusive dynamical laws~\cite{granger69,pearl00,runge18}. 
Efforts to understand those laws have prompted considerable advances in network science during the last few decades.
The journey started with models for complex networks with a single type of dyadic links \cite{albert02,latora17,newman18}, continued with multilayer and multiplex networks \cite{kivela14,boccaletti14}, and has most recently brought us different approaches to model higher-order interactions in complex systems \cite{berge84,hatcher02,holme15,benson16,lambiotte19,fede20,torres20,bick23,fede21}. 
Over time, we have learned to impose weaker assumptions upon our models and thus to describe more complex interactions.

Lately, the community has focused on the study of two types of interactions: multibody interactions and time-ordered interactions.
The former are modeled with hypergraphs and the latter are modeled with higher-order Markov chains.
Each of them violates a different assumption of standard network models:
multibody interactions violate the assumption that system dynamics can be explained solely with dyadic interactions;
time-ordered interactions violate the Markov assumption \cite{markov06,cover06}.
However, real systems may violate both assumptions, invalidating either modeling approach. 
This situation leads to multiple open questions.
How can we analyze systems that exhibit interactions that are both time-ordered and multibody?
More specifically, how can we formalize such interactions and infer them from data?
How can we use time-ordered multibody interactions to explain system dynamics?

The multibody facet of the problem has already been solved with the inference of multibody interactions from dynamical equations \cite{casadiego17} and with information theory tools \cite{williams10,rosas19,rosas20}. 
There are also preliminary efforts to unify the modelling of time-ordered and multibody interactions, e.g. in the study of synchronization \cite{zhang20}, contagion \cite{chowdary21} and consensus dynamics \cite{leonie21}. 
However, as discussed in \cite{fede21} a general formalism is yet to be found.

In this manuscript, we propose a unified methodology for scenarios in which multibody and time-ordered dependencies coexist.
First, we introduce parsimonious models of time-ordered multibody interactions.
Then, we present an algorithm to decompose the system dynamics into an ensemble of interactions.
Last, we introduce a measure to characterise the complexity of such ensembles of interactions.
We show how the integration of time-ordered and multibody interactions enables a better description of real world systems than current modeling paradigms.

\begin{figure*}[t]
\captionsetup[subfigure]{labelformat=empty}
\centering
\subfloat[]{\includegraphics[width=\textwidth]{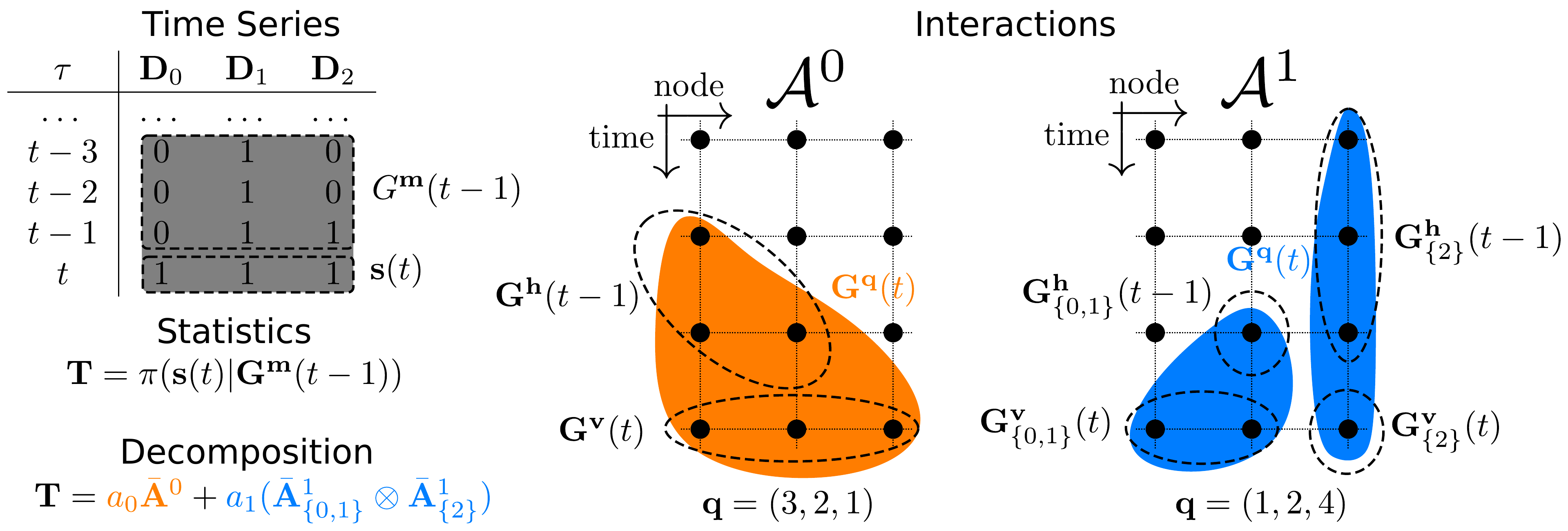}}
\caption{We are presented the task to explain the dynamics of a system of $N=3$ nodes from their temporal data $\mathbf{D}$.
We first compute a higher-order Markov chain of memory $m=3$ and collect the transition probabilities in a matrix $\T$, which predicts the joint node state $\mathbf{s}(t)$ as a function of the previous three states $\G^{\mathbf{m}}(t-1)$. 
We then decompose $\T$ as a convex sum of time-ordered multibody interactions $\interaction^i$.
An interaction of type $\q$ models the dynamics as a function of the group of system variables $\G^{\mathbf{q}}(t)$, which contains $q_n$ variables of each node $n$. 
$\G^{\mathbf{q}}(t)$ is divided in $\G^{\mathbf{v}}(t)$ and $\G^{\mathbf{h}}(t-1)$ that account respectively for row and column variables of interaction transition matrices $\bar{\interactionM}^i$.
The interaction $\interaction^1$ is separated into two dynamically independent subsystem interactions $\interaction^1_{\{0,1\}}$ and $\interaction^1_{\{2\}}$.} 
\label{fig:protocol}
\end{figure*}

\section{Time-ordered multibody interactions}
\subsection{Notation}
Let $S$ be a closed system of $N$ nodes $n\in \mathcal{N}$, $\mathcal{N}=\{n\}^{N-1}_{0}$.
Each node $n$ is constrained to a finite set of states, called alphabet $X_n$  with cardinality $|X_n|$.
We denote the state of each node $n$ at a given time $t$ with $s_n(t)$, and the state of the whole system with $\mathbf{s}(t) = (s_n(t))_{n\in\mathcal{N}}$. 
Thus, the alphabet of the whole system is $X=\times_{n \in \mathcal{N}} X_n$, where $\times$ denotes Cartesian product.
We observe the system in an interval  $(t_0, t_\text{end})$ and collect sequences of system states $\mathbf{D}=(\mathbf{s}(\tau))^{t_{\textrm{end}}}_{\tau = t_{0}}$ and element states $\mathbf{D}_n=(s_n(\tau))^{t_{\textrm{end}}}_{\tau = t_{0}}$. 

We can model the dynamics of $S$ as a higher-order Markov chain with memory $m$: the transition probability to a state $x \in X$ depends on the previous $m$ states $\bar{x} \in |X|^{m}$.
The transition probabilities $\pi(x|\bar{x})$ are encoded in matrix elements $T_{x\bar{x}}$ of transition matrices $\T$ with dimensions $|X| \times |X|^{m}$.
One can infer the transition matrix $\T$ for different Markov orders $m$ using the temporal data $\mathbf{D}$, and select the Markov order that provides the best predictability-parsimony balance \cite{aic74,bic78,scholtes17}.
See Fig. \ref{fig:protocol} for an example of our notation.

Higher-order Markov chains model the dynamics of $S$ as a function of all nodes at all times and thus one cannot analyze node interdependencies in isolation. 
Therefore, we deviate from this modelling approach, and define time-ordered multibody interactions that depend on the last $q_n$ states of every node $n$.
We denote a group of $q_n$ successive temporal states of node $n$ with tuples $\G_n^{q_n}(t):=(s_n(\tau))^{t}_{\tau=t-q_n+1}$.
If $q_n$ is zero, $\G_n^{q_n}(t)$ denotes an empty tuple.
We then construct a vector $\mathbf{q}=(q_n)_{n \in \mathcal{N}}$ and denote groups of variables for the whole system $\G^\mathbf{q}(t):=(\G_n^{q_n}(t))_{n \in \mathcal{N}}$. States of $\G^\mathbf{q}(t)$ take values in alphabet $X^{\mathbf{q}} := \times_{n \in \mathcal{N}} X_n^{q_n}$.
Before defining time-ordered multibody interactions, we emphasize that in our notation Markov chains with memory $m$ model $\mathbf{s}(t) = \G^\mathbf{1}(t)$ given the history $\G^\mathbf{m}(t-1)$, where $\mathbf{1}$ and $\mathbf{m}$ are vectors with all components $1$ and $m$, respectively. 
The transition probabilities then depend on $\order := m+1$ system states, amounting to a total $\order N$ time variables $\G^\mathbf{\order}(t)$, where $\mathbf{\order}$ is a vector with all components $\order$. See Fig. \ref{fig:protocol}.

\subsection{Interactions}
The core of our formalism are time-ordered multibody interactions, which depend only on a subset of the $\G^{\mathbf{\order}}(t)$ variables of higher-order Markov chains. 
A time-ordered multibody interaction $\interaction$ of type $\mathbf{q}$, $0\leq q_n \leq \order$, models the probability of $\mathbf{s}(t)$ given history $\G^\mathbf{m}(t-1)$ as a function of variables $\G^\mathbf{q}(t)$ and independent of other variables. 
We separate the variables $\G^\mathbf{q}(t)$ in two groups: variables $\G^\mathbf{v}(t)$ describing nodes at time $t$ and variables $\G^\mathbf{h}(t-1)$ describing their history. 
Here, the components of $\mathbf{v}$ are $v_n=0$ when $q_n=0$, and $v_n=1$ when $q_n>0$; the vector $\mathbf{h}$ contains history lengths $\mathbf{h} = \mathbf{q} - \mathbf{v}$.
See Fig. \ref{fig:protocol} for an example of an interaction ensemble with two interactions in a system of three nodes, and corresponding $\G^\mathbf{v}(t)$ and $\G^\mathbf{h}(t-1)$.
The variables $\G^\mathbf{v}(t)$ are modeled with a multinomial distribution that depends on the history $\G^\mathbf{h}(t-1)$.
The state of $\G^\mathbf{v}(t)$ takes values $y \in X^\mathbf{v}$, and the state of $\G^\mathbf{h}(t-1)$ takes values $\bar{y} \in X^\mathbf{h}$.
The transition probabilities from $\bar{y}$ to $y$ are encoded in the transition matrix $\interactionMel_{y\bar{y}}$ of interaction $\interaction$.
For nodes where $q_n = 0$, the probabilities of $s_n(t)$ do not depend on any variable, and thus they are uniform.
In summary, an interaction $\interaction$ models the dynamics as:
\begin{align}
    \pi(\mathbf{s}(t)| \G^\mathbf{m}(t-1), \interaction) = 
    \frac{\interactionMel_{y\bar{y}}}{\prod_{n \in  \mathcal{N} | q_n = 0}|X_n|}.
    \label{eq:interaction}
\end{align}
While the transition matrix $\T$ has $|X| \cdot |X|^m = \prod_{n} |X_n|^\order$ elements, the interaction $\interaction$ of type $\mathbf{q}$ has only $|X^\mathbf{v}| \cdot |X^\mathbf{h}| = \prod_{n \in \mathcal{N}} |X_n|^{q_n}$ parameters. 
Therefore, interactions produce a parsimonious model of the system dynamics.
With this, we have formally introduced time-ordered multibody interactions, and we now discuss how we can use them to decompose the dynamics of $S$.

\subsection{Interaction hierarchy}
\begin{figure}[t]
\centering
\includegraphics[width=1\columnwidth]{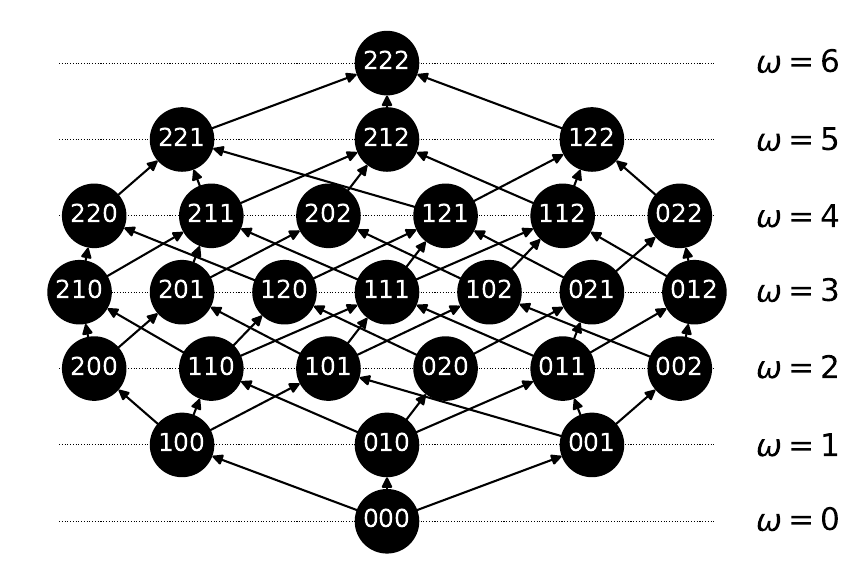}
\caption{Hasse diagram of time-ordered multibody interactions for a system of three nodes and memory equal to one.
Interaction types are depicted as vertices $\mathbf{q}$, and arranged according to their interaction order $\omega$. 
If there is a path from $\mathbf{q}$ to $\mathbf{q}'$, then every interaction of type $\mathbf{q}$ can be expressed as an interaction of type $\mathbf{q}'$.}
\label{fig:hasse}
\end{figure}

The aim of our decomposition is to explain the dynamics in terms of the simplest possible interactions. 
Therefore, we organize time-ordered multibody interactions in a hierarchy by their complexity.
We define the interaction order $\omega$ of interaction type $\mathbf{q}$ as $\omega=\sum_{n \in \mathcal{N}} q_n$, that counts the number of variables on which the interaction depends.
Moreover, interactions are nested models: an interaction $\interaction$ of type $\mathbf{q}$ can be nested in (or represented with) an interaction $\interaction'$ of type $\mathbf{q}'$ iff $\forall n: q_n \leq q'_n$.
Intuitively, if variables of $\interaction$ are a subset of variables of $\interaction'$, $\interaction$ can be nested in $\interaction'$. 
The relation of nestedness over the set of interaction types naturally defines a partial order, which is commonly depicted with a Hasse diagram.
In the Hasse diagram on Fig \ref{fig:hasse}, two interaction types are connected as long as they are nested and separated by one interaction order $\omega$.
For example, an interaction of type $\mathbf{q}=(2,1,0)$ (leftmost type in fourth row of Fig. \ref{fig:hasse}) models the dynamics of $S$ as a function of $s_0(t)$, $s_0(t-1)$, and $s_1(t)$. 
Thus, it can be nested in any interaction type that includes these variables.
The interaction type $\mathbf{q}=\mathbf{\order}$, which corresponds to a Markov chain on $S$, is particularly important for the decomposition because any other type of interaction can be nested in it.
When we nest an interaction $\interaction$ in type $\mathbf{\order}$, we denote the resulting interaction {with $\bar\interaction$ and its transition} matrix with $\bar\interactionM$.

\subsection{Decomposition algorithm}

The decomposition algorithm has two steps: in a first step we decompose the transition matrix $\T$ into an ensemble of time-ordered interactions; in a second step, we decompose the time-ordered interactions into subsystem interactions \footnote{Code available at \url{https://github.com/unaialro/inin}}.
Formally, in the first part we find the interaction coefficients $a_i$ and matrices $\interactionM^i$ (here, $i$ is an upper index) such that the matrix $\T$ can be expanded as a convex sum: $\T=\sum_i a_i \bar\interactionM^i$.
The algorithm repeatedly extracts interactions from $\T$.
Starting at $\omega=0$, it explores all interactions at $\omega$ before proceeding with $\omega +1$ and extracts the one with a highest coefficient.
Let us assume we have already extracted $i$ interactions from $\T$ and that $\tilde\T$ is the remainder: $\tilde\T=\T - \sum_{j=0}^{i-1} a_j \bar\interactionM^j$. 
For the next interaction $\interaction^i$, the algorithm finds the transition matrix $\interactionM^i$ such that the corresponding coefficient $a_i$ is maximal and $\tilde\T - a_i \bar\interactionM^i$ does not have negative elements.
The first part finishes when either $\tilde\T$ is a matrix of zeros, or, equivalently, when the sum of the interaction coefficients is one.
See Appendix \ref{ap:subin} for the details.

The second step of the algorithm identifies statistically independent subsystems and decomposes the interactions into subsystem interactions.
Formally, subsystem interactions are interactions defined on a subset of nodes.
Let $\mathrm{P}(\mathcal{N})$ be a partition of $\mathcal{N}$, and $\mathcal{B}\in \mathrm{P}(\mathcal{N})$ be a subset of nodes $\mathcal{B}\subset\mathcal{N}$.
If $x$ is a state of nodes $\mathcal{N}$ and $\bar{x}$ is the history of the state, we denote the corresponding state of the nodes $\mathcal{B}$ with $z_\mathcal{B}$ and their corresponding history $\bar{z}_\mathcal{B}$.
We say that interaction $\interaction$ has a subsystem interaction partition $\mathrm{P}(\mathcal{N})$ iff
\begin{equation}
\pi(x|\bar{x}, \interaction)=
\prod_{\mathcal{B} \in \mathrm{P}(\mathcal{N})} \pi(z_\mathcal{B}|\bar{z}_\mathcal{B}, \interaction_{\mathcal{B}}).
\end{equation}
For instance, nodes with $q_n~=~0$ are obvious examples of independent subsystem interactions.

We identify subsystem interactions from an interaction $\interaction$ by iteratively factorizing its transition matrix as a tensor product.
First, we factor out each node $n$ that is modeled with a uniform distribution $q_n = 0$.
We thus obtain the partition with one-element subsets for nodes $n$ with $q_n =0$, and a single subset containing the remaining nodes. 
Then, we iteratively search for two-group partitions of the subsets with more than one element.
We stop when we cannot find a valid partition of any subset.
See Appendix \ref{ap:subin} for the details on the subsystem interaction factorization, and see SM for a selection of practical examples of the full protocol including the Glauber dynamics \cite{glauber}.

\subsection{Complexity}
Having discussed the formalism of time-ordered multibody interactions, and the algorithm to extract them from data capturing the system-level dynamics of node states, we now focus on the complexity of interaction ensembles.
We note that the expansion of the transition matrix $\T$ is not unique, as it can be reconstructed with different ensembles of interactions and interaction coefficients.
Considering Occam's razor, here we are interested in an ensemble with the lowest possible complexity.
We thus need a principled way of measuring the complexity for different combinations of interactions and subsystem interactions. 

\begin{figure}[t!]
\centering
\subfloat[]{\includegraphics[width=0.24\textwidth]{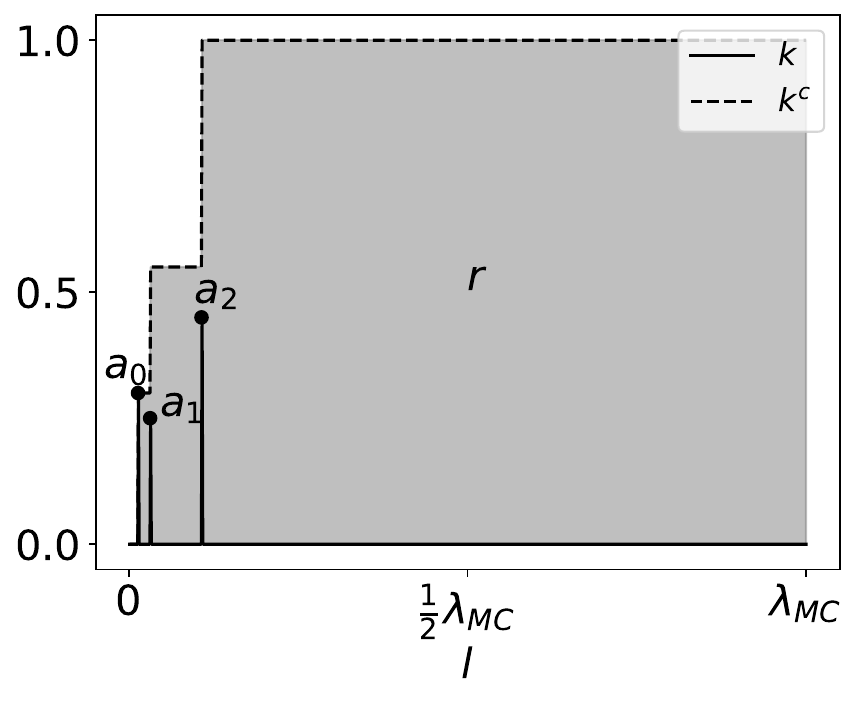}}
\subfloat[]{\includegraphics[width=0.24\textwidth]{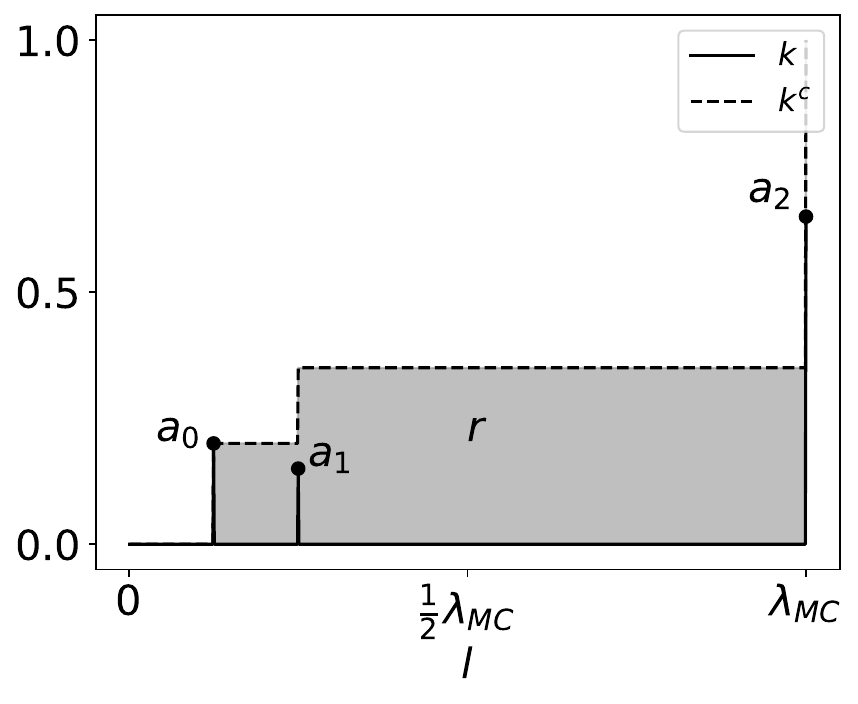}}\\[-0.25cm]
\caption{Complexity measures $k(l)$, $k^c(l)$ and $r$ for (a) high reducibility and (b) low reducibility systems.
Here $\lambda(\interaction)$ is the number of free parameters of an interaction, $k(l)$ weights how likely is to have an interaction with $l$ parameters in an interaction ensemble, and $k^c(l)$ is the cumulative of $k(l)$.
The reducibility $r$ is the share of area under $k^c(l)$, and encodes the relative reduction in the average number of parameters per interaction of an ensemble.
The ensemble in (a) has the same predictive power as $\T$ even if its matrices contain $94 \%$ less parameters on average.}
\label{fig:redu} 
\end{figure}

To quantify the complexity of an ensemble of interactions, we introduce a the measure of reducibility.
Let us denote the number of degrees of freedom of an interaction $\interaction$ with $\lambda(\interaction)$.
When interactions have no subsystem interactions, $\lambda(\interaction)$ is simply $(|X^{\mathbf{v}}|-1)|X^{\mathbf{h}}|$ due to the normalization of stochastic matrices.
If $\interaction$ has subsystem interactions, $\lambda(\interaction)$ is given by the sum of their parameters $\lambda(\interaction)=\sum_{\mathcal{B}} \lambda(\interaction_{\mathcal{B}})$.
We call the ensemble reducible if simple interactions have a high contribution in the interaction ensemble. We call it irreducible if complex interactions have a high contribution in the interaction ensemble.
Formally, reducibility $r$ is defined as:
\begin{equation}
r
= \frac{\lambda_\text{MC} - \sum_i a_i \lambda(\interaction^i)}
{\lambda_\text{MC}},
\end{equation}
where $\lambda_\text{MC} = (|X|-1)|X|^m$ is the number of degrees of freedom of the Markov chain.
Therefore, $r$ accounts for the reduction in the average degrees of freedom when our interaction ensembles are used instead of Markov chains.

We showcase the power of this measure in the context of lossy compression. 
Let us denote with $k(l)$ the probability mass function of all interactions with $l$ degrees of freedom: $k(l)=\sum_{\lambda(\interaction^i) = l} a_i$, and its cumulative distribution with with $k^c(l)$. 
The value of $k^c(l)$ describes how well the interactions with less than $l$ degrees of freedom represent the system dynamics.
In Fig. \ref{fig:redu}, we depict $k(l)$ and $k^c(l)$ of a reducible and irreducible interaction ensembles.
When the ensemble is reducible (panel (a)), it is dominated by the simple interactions, thus $k(l)$ peaks for small values of $l$, and $k^c(l)$ is high throughout the interval.
When the ensemble is irreducible (panel (b)), $k(l)$ has peaks for large values of $l$, and $k^c(l)$ is low throughout the interval.
Therefore, when an ensemble is reducible, we can choose a small $l$ where $k^c(l)$ is high.
Thus, by only using the interactions with less than $l$ degrees of freedom, we can compress a reducible ensemble with an approximate loss of $1-k^c(l)$, as this accounts for the interactions that are above $\lambda=l$.
Lastly, it is worth noting that the reducibility is the share of the area under $k^c(l)$. 

\begin{table}[b] 
\centering
\renewcommand{\arraystretch}{1.2}
\begin{tabular}{c|ccccccc}
\hline
label & $q$ & $\mathcal{B}$ & $\omega$ & $|X^{\mathbf{v}}|$ & $|X^{\mathbf{h}}|$ & $\lambda(\interaction)$ & $a$ \\
\hline
$\interaction^0$ & $210$ & $\{0,1,2\}$ & $3$ & $4$ & $2$ & $6$ & $0.3$ \\
$\interaction^1$ & $132$ &
\begin{tabular}{c}
$\{0,1\}$ \\ $\{2\}$
\end{tabular}
&
\begin{tabular}{c}
$4$\\$2$
\end{tabular}
&
\begin{tabular}{c}
$4$\\$2$
\end{tabular}
&
\begin{tabular}{c}
$4$\\$2$
\end{tabular}
&
\begin{tabular}{c}
$12$\\$2$
\end{tabular}
& 0.25 \\
$\interaction^2$ & $033$ & $\{0,1,2\}$ & $6$ & $4$ & $16$ & $48$ & $0.45$\\
\hline
\end{tabular}
\caption{Interaction type $\q$, subsystem interaction partition $\mathcal{B}$, interaction order $\omega$, matrix rows $|X^{\mathbf{v}}|$ and columns $|X^{\mathbf{h}}|$, number of parameters $\lambda(\interaction)$, and coefficient $a$ for the interaction ensemble of Fig. \ref{fig:redu} a.}
\label{tab:redu}
\end{table}

\section{Validation}
Finally, we present a series of experiments to validate different aspects of our contributions.
In a first experiment, we establish a relation between the accuracy of our algorithm and the available amount of data. Our goal here is to understand how an error in $\T$ propagates to errors in $a$ and $r$. 
We randomly generate $\T$ by drawing its elements from a uniform distribution and obtain the ground truth values of interaction coefficients $\mathbf{a}$ and reducibility $r$.
Then, we construct random sequences out of $\T$, analyse them with our algorithm, and obtain estimators $\hat{\T}$, $\hat{\mathbf{a}}$, $\hat{r}$.
We evaluate the accuracy of all three estimators in terms of the total variation distance $\sigma$ between the ground truth and the estimated values.

\begin{figure}[t]
\centering
\subfloat[]{\includegraphics[width=0.2\textwidth]{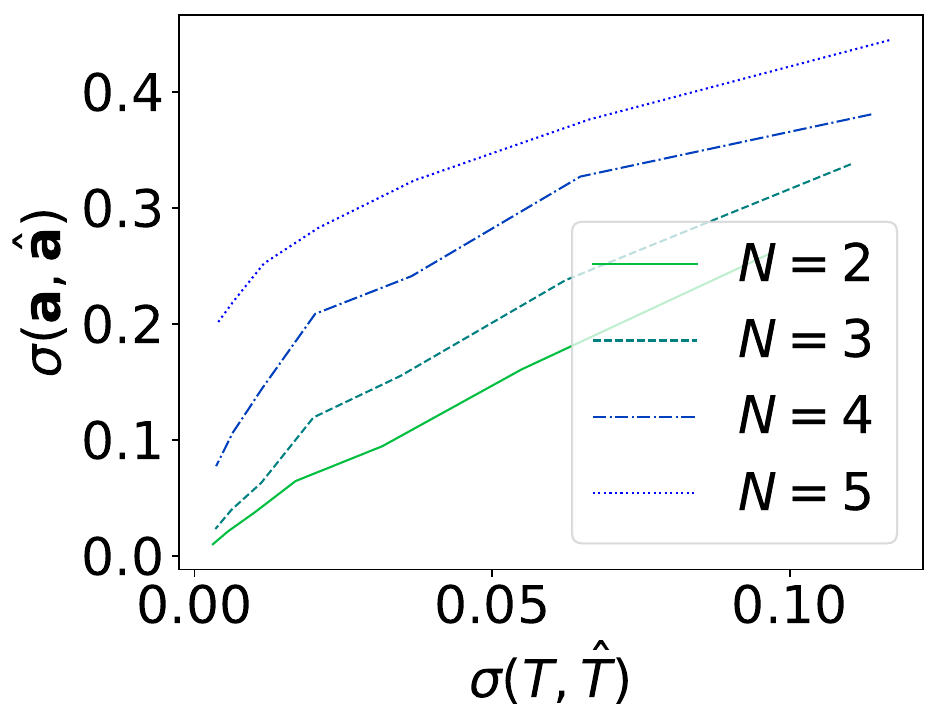}}
\subfloat[]{\includegraphics[width=0.2\textwidth]{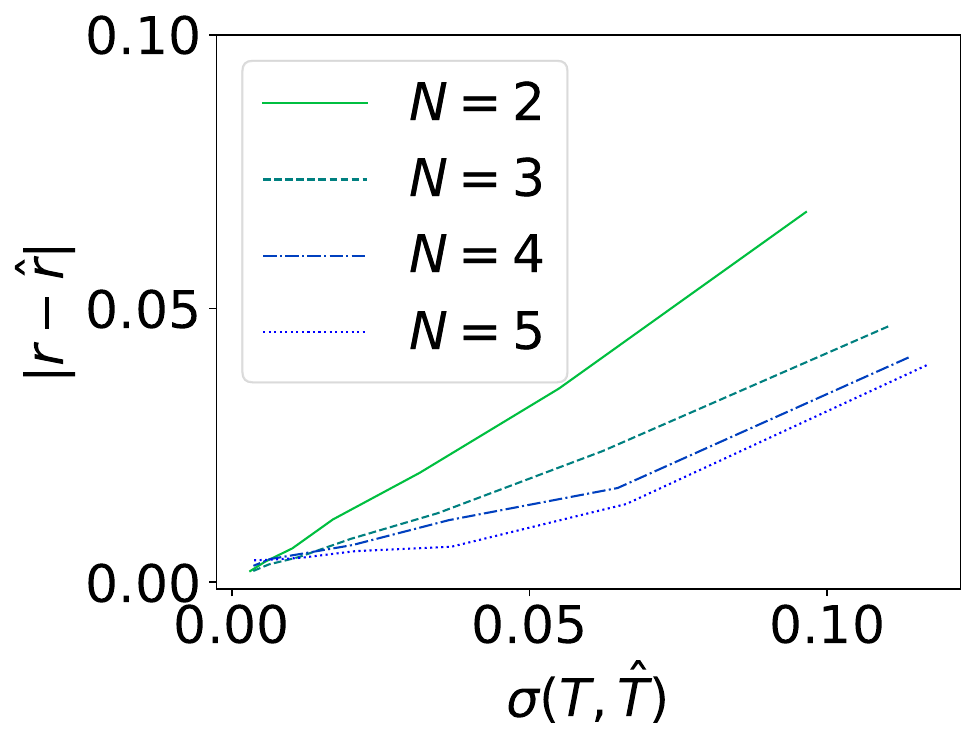}}\\[-0.25cm]
\caption{We produce synthetic sequences and compute the statistical distance between the original transition matrices $\T$, (a) interactions $\mathbf{a}$ and (b) reducibilities $r$, and the ones inferred from the data $\hat{\T}, \hat{\mathbf{a}}, \hat{r}$ for a fixed alphabet $|X_n|=2$ and order $o=2$. Results reported here correspond to averages over $100$ realizations.} 
\label{fig:val}
\end{figure}

The results are reported in Fig. \ref{fig:val}, where we explore how the errors dependend on the system size $N$ for a system with all equal alphabets $|X_n|=2$ and order $\order=2$ (see SM for a more general selection of parameters).

As noted above, given a dataset there is not a unique interaction ensemble. This explains why in Fig. \ref{fig:val}a, inferred interactions deviate from the ground truth. However, and despite errors in inferring interactions, Fig. \ref{fig:val}b reveals the capacity of the algorithm to explain the dynamics with an ensemble whose reducibility is close to the ground truth value. Therefore, the protocol is valid for obtaining simple and accurate interaction ensembles.   

We perform a second experiment to evaluate the capacity of our algorithm to find high-reducibility decompositions.
For this, we compare our protocol with an alternative decomposition algorithm that we call random algorithm (RA).
In RA interactions are randomly selected from the interaction diagram. 
The experiment, Table \ref{tab:rand}, shows that our algorithm consistently outperforms the random algorithm RA, and therefore validates our algorithm for inferring time-ordered multibody interactions for the first time.

\begin{table}[t] 
\centering
\renewcommand{\arraystretch}{1.2}
\begin{tabular}{c|*{7}{p{0.1\columnwidth}}}
\hline
$(\order, N)$ & $(2,2)$ & $(2,3)$ & $(2,4)$ & $(2,5)$ & $(3,2)$ & $(3,3)$ & $(3,4)$ \\
\hline
$r_{RA}$ & 0.35 & 0.38 & 0.42 & 0.46 & 0.30 & 0.33 & 0.38  \\
$r$ & 0.55 & 0.56 & 0.58 & 0.62 & 0.44 & 0.46 & 0.51 \\
\hline
\end{tabular}
\caption{Average values of reducibilities $r_{RA}$ computed with a random algorithm and $r$ computed with our method, over $1000$ realizations. We decomposed the transition matrices of systems with parameters $\order$, $N$ and $|X_n|=2$.}
\label{tab:rand}
\end{table}

\section{Conclusion}
In summary, we developed a framework for studying complex systems whose dynamics is ruled by time-ordered and multibody dependencies. 
The framework is based on interactions, which emerge from the expansion of time-evolution operators for multivariate higher-order Markov chains.
We proposed a measure for the complexity of interaction ensembles and an algorithm to extract them from time series data.
We believe these contributions to be a relevant asset for the field of complex systems as they address a currently latent problem by reconciling a data-oriented perspective with an analytical description of node interdependencies.

\section{Acknowledgments}
The authors acknowledge support from the Swiss National Science Foundation (SNSF) via grant number 176938. We also thank Vincenzo Perri for useful comments.

\bibliography{/home/unai/Documents/tr/ref/bio.bib}

\newpage

\section{Appendix}
\subsection{Higher-order Markov chains}
\label{ap:homc}
Higher-order Markov chains will be able to describe the dynamics of $S$ as long as the following assumptions hold: (1) The same statistical behaviour is expected at all points in time (causal stationarity). (2) No external variables can influence the dynamics of $S$ (causal sufficiency) \cite{runge18}.

\subsection{Nested interactions}
An interaction $\interaction$ of type $\q$ is nested in an interaction $\interaction'$ of type $\q'$ by fixing the transition probabilities of $\interaction'$ to those of $\interaction$:
\begin{equation}
\pi(\G^{\mathbf{v}'}(t)|G^{\mathbf{h}'}(t-1))=\frac{\pi(\G^{\mathbf{v}}(t)|\G^{\mathbf{h}}(t-1))}{\prod_{\substack{q'_n\neq0 \\ q_n = 0}} |X_n|}.
\label{eq:enlarge}
\end{equation}

\subsection{Algorithm}
\label{ap:alg}
We now explain the procedure to extract an interaction at interaction order $\omega$. 
This procedure has been designed for nodes with all equal alphabets $X_n=X_{n'}$ $\forall$ $n,n' \in \mathcal{N}$.
If nodes were different our algorithm would also yield a valid decomposition, but it should be modified for improved results.
\\

At each round of the algorithm we extract the interaction $\interaction$ with the highest coefficient $a$ for a given interaction order $\omega$.
Therefore, we first explore all interactions $\interaction$ of interaction order $\omega$ to obtain their interaction matrix $\interactionM$. This matrix, models transitions from histories $\bar{y} \in X^{\mathbf{h}}$ to states $y \in X^{\mathbf{v}}$.

Let us explain how to obtain an interaction matrix $\interactionM$ from $\tilde{\T}$:  
For every $y \in X^{\mathbf{v}}$ and $\bar{y} \in X^{\mathbf{h}}$ we create a set $\mathcal{E}$ with matrix elements of $\tilde{\T}_{x\bar{x}}$. In order to decide which elements of $\tilde{\T}_{x\bar{x}}$ are in $\mathcal{E}$ we define two functions $F_{y}(x)$ and $F_{\bar{y}}(\bar{x})$:
\begin{equation}
F_{y}(x): X \rightarrow \{\textrm{True,False}\}, \hspace{0.25cm} F_{\bar{y}}(\bar{x}): X^m \rightarrow \{\textrm{True,False}\},
\label{sm_eq:funct}
\end{equation}
which characterize respectively the rows $x$ and columns $\bar{x}$ of $\tilde{\T}$ that should be included in $\mathcal{E}$. 

\subparagraph{$F_{y}(x)$} The function is computed as follows:
\begin{algorithmic}[1]
\Function{$F_{y}$}{$x$}
    \For{$n \in \mathcal{N}$}
        \If{$y_n \neq \emptyset$ $\land$ $y_n \neq x_n$}
            \State \Return{False}
        \EndIf
    \EndFor
    \State \Return{True}
\EndFunction
\end{algorithmic}
If the component of node $n$ of $x$ and $y$ coincide for all nodes in $\mathcal{N}$, with the exception of nodes not in $y$, the row has to be included in $\mathcal{E}$. If for any node the components of $x$ and $y$ are different, the function will output ``False" before finishing the loop.  

\subparagraph{$F_{\bar{y}}(\bar{x})$} The function is computed as follows:
\begin{algorithmic}[1]
\Function{$F_{\bar{y}}$}{$\bar{x}$}
    \For{$n \in \mathcal{N}$}
        \For{$k \in \{0,...,m-1\} $}
            \If{$\bar{y}_{nk} \neq \emptyset$ $\land$ $\bar{y}_{nk} \neq \bar{x}_{nk}$}
                \State \Return{False}
            \EndIf
        \EndFor
    \EndFor
    \State \Return{True}
\EndFunction
\end{algorithmic}
In this case $\bar{y}$ and $\bar{x}$ are bidimensional tuples. A first dimension accounts for the node $n$, and the second dimension accounts for the memory. The idea is the same, we check coincidence for all nodes and for all memories between $\bar{x}$ and $\bar{y}$, with the exception of nodes that are not in $\bar{y}$, or nodes whose memory is shorter than $m-1$.

$\mathcal{E}$ contains those elements $\tilde{T}_{x\bar{x}}$ that fulfill $F_{y}(x)$ and $F_{\bar{y}}(\bar{x})$. The minimum of such set is the element $R_{y\bar{y}}$ of the auxiliary matrix $\mathbf{R}$:
\begin{equation}
R_{y\bar{y}}~=~\min \mathcal{E}.
\end{equation}
This process is repeated for all $y$ and $\bar{y}$, to complete $\R$. We then obtain $\interactionM$ by normalizing $\mathbf{R}$ column-wise, i.e.,
\begin{equation}
\interactionM_{\bar{y}}=\frac{\R_{\bar{y}}}{||\R_{y}||_1},
\end{equation}
where $\interactionM_{\bar{y}}$ are the columns of $\interactionM$ and $\R_{\bar{y}}$ are the columns of $\R$.

Once we have obtained all possible interactions at order $\omega$, we represent them as interactions of type $\mathbf{\order}$, and the transition matrices $\bar\interactionM$ have the same dimensions as $\tilde{\T}$.
For each interaction we find the maximal value of $a$ such that all elements of  $\tilde{\T}-a\bar\interactionM$ are non-negative: we compute minimal column sum of $\mathbf{R}$, and include the uniform distribution for the states of nodes $n$ for which $q_n=0$. The coefficients read
\begin{equation}
a_i=\left( \min_{\bar{y}} \sum_y R_{y \bar{y}} \right) \prod_{q_n=0} |X_n|.
\label{eq:coeff}
\end{equation}%
When all interactions at interaction order $\omega$ yield $a=0$ we explore the next interaction order $\omega+1$.

In the third step we select the interaction with the highest interaction coefficient $a$, and we extract it from $\tilde{\T}$: the new value, $\tilde{\T'}$, is given by $\tilde{\T'} = \tilde{\T} - a \bar\interactionM$.
This sequence of steps is repeated until $\tilde{\T'}$ is a matrix of zeros, or equivalently until $\sum_i a_i = 1$. See the next subsection for an example.

\subsection{Decomposition example}
Consider the following example of $\tilde\T$ in a system with $N=2$, $\order=2$ and $|X_n|=2$.
We here show an iteration of the decomposition algorithm step by step.
The interaction to be extracted will be interaction $\interaction$ with type $\mathbf{q}=(0,2)$. Therefore we have $\G^{\mathbf{o}}(t)=\G^{(2,2)}(t)$ and $\G^\q(t)=\G^{(0,2)}(t)$ to characterize the complete system variables and the interaction variables respectively. The auxiliary matrix $\mathbf{R}$ is obtained from $\tilde\T$ as
\begin{equation}
\tilde{\T}=\frac{1}{100}\left( \begin{array}{cccc} \dda{33} & \ddb{22} & \dda{13} & \ddb{64} \\ \ddc{3} & \ddd{34} & \ddc{4} & \ddd{14} \\ \dda{47} & \ddb{2} & \dda{23} & \ddb{17} \\ \ddc{17} & \ddd{42} & \ddc{60} & \ddd{5} \end{array}  \right) \longrightarrow 
R= \frac{1}{100}\left( \begin{array}{cc} \dda{13}& \ddb{2} \\ \ddc{3} & \ddd{5} \end{array}  \right).
\label{eq:red}
\end{equation} 
Let us visualize how $R_{00}$ has been computed. $R_{00}$ accounts for $\pi(\G^{(0,1)}(t)=0|\G^{(0,1)}(t-1)=0)$, and is thus associated with the elements of $\tilde\T$ that encode the same transition probability complemented with all the possible values for $\G^{(1,0)}(t)$ and $\G^{(1,0)}(t-1)$:
\begin{eqnarray*}
\pi(\G^{(1,1)}(t)=(0,0)|\G^{(1,1)}(t-1)=(0,0))=33/100, \\
\pi(\G^{(1,1)}(t)=(0,0)|\G^{(1,1)}(t-1)=(1,0))=13/100, \\
\pi(\G^{(1,1)}(t)=(1,0)|\G^{(1,1)}(t-1)=(0,0))=47/100, \\
\pi(\G^{(1,1)}(t)=(1,0)|\G^{(1,1)}(t-1)=(1,0))=23/100. \\
\end{eqnarray*}
The auxiliary matrix is constructed with the minimal of those elements such that when $\interactionM$ is removed from $\tilde\T$ no negative elements are created in the next iteration.\\

The second step yields $a=(2/100 + 5/100) \cdot 2$, where the first part comes from $R_{01}+R_{11}$, and the second part comes from the dimensions of $|X_0|=2$,
\begin{equation}
\mathbf{R}=\frac{1}{100}\left( \begin{array}{cc} 13 & 2 \\ 3 & 5 \end{array}  \right) \longrightarrow \interactionM=\left( \begin{array}{cc} \frac{13}{16} & \frac{2}{7} \\ \frac{3}{16} & \frac{5}{7} \end{array}  \right).
\end{equation}

In the third step we move from a $\mathbf{q}=(0,2)$ representation of $\interactionM$ to a $\mathbf{q}=(2,2)$ representation:
\begin{equation}
\left( \begin{array}{cc} \dda{\frac{13}{16}} & \ddb{\frac{2}{7}} \\ \ddc{\frac{3}{16}} & \ddd{\frac{5}{7}} \end{array}  \right) \longrightarrow \left( \begin{array}{cccc} \dda{\frac{13}{2 \cdot 16}} & \ddb{\frac{2}{2 \cdot 7}} & \dda{\frac{13}{2 \cdot 16}} & \ddb{\frac{2}{2 \cdot 7}} \\ \ddc{\frac{3}{2 \cdot 16}} & \ddd{\frac{5}{2 \cdot 7}} & \ddc{\frac{3}{2 \cdot 16}} & \ddd{\frac{5}{2 \cdot 7}} \\ \dda{\frac{13}{2 \cdot 16}} & \ddb{\frac{2}{2 \cdot 7}} & \dda{\frac{13}{2 \cdot 16}} & \ddb{\frac{2}{2 \cdot 7}} \\ \ddc{\frac{3}{2 \cdot 16}} & \ddd{\frac{5}{2 \cdot 7}} & \ddc{\frac{3}{2 \cdot 16}} & \ddd{\frac{5}{2 \cdot 7}} \end{array} \right).
\end{equation}  
At this point it becomes clear that the correction in the dimension at Appendix Eq. \eqref{eq:coeff} is necessary to normalize the $q_n=\order$ form of $\interactionM$ to model nodes with $q_n=0$.

\subsection{Subsystem Interactions}
\label{ap:subin}
We organize the transition probabilities of the model in a transition matrix $\mathbf{W}$ ordered according to $\mathcal{B}$.
The output matrices $\mathbf{U}$ and $\mathbf{V}$, with dimensions $u_r \times u_c$ and $v_r \times v_c$ read
\begin{equation}
U_{ij}= \sum^{v_c -1}_{k=0} W_{i v_r +k,j v_c}, \hspace{0.5cm} V_{ij}= \sum^{u_c -1}_{k=0} W_{k v_r + i,j}.
\label{sm_eq:sub}
\end{equation}
If $\mathbf{W}=\mathbf{U}\otimes \mathbf{V}$ holds $\mathbf{U}$ and $\mathbf{V}$ are valid subsystem interaction matrices.
In that case, the procedure should be applied again on $\mathbf{U}$ and $\mathbf{V}$ until no more subsystems can be isolated.
For an interaction $\interaction$ of type $\q$ subsystem interactions at nodes $\mathcal{B}$ have $\prod_{n\in \mathcal{B}} |X_n|^{q_n}$ parameters, and the decomposed interaction has only $\sum_{\mathcal{B}} \prod_{n\in \mathcal{B}} |X_n|^{q_n}$ degrees of freedom.

Let $\interactionM$ be the transition matrix of an interaction with $N=2$, $|X_0|=3$ and $|X_1|=2$. 
Our goal here is to decompose $\interactionM$ as $\interactionM=\mathbf{U} \otimes \mathbf{V}$ with $\mathbf{U} \in \mathbb{R}^{3 \times 3}$ and $\mathbf{V} \in \mathbb{R}^{2 \times 2}$.
If we assumed $\mathbf{U} \otimes \mathbf{V}$ the following would hold:
\begin{align}
\nonumber &\left( \begin{array}{*{3}{c}} u_{00}&u_{01}&u_{02}\\u_{10}&u_{11}&u_{12}\\u_{20}&u_{21}&u_{22}\\ \end{array} \right) \otimes \left( \begin{array}{*{2}{c}} v_{00}&v_{01}\\v_{10}&v_{11} \end{array} \right) =  \\
&\left( \begin{array}{cc|cc|cc} 
    u_{00}v_{00}&u_{00}v_{01}&u_{01}v_{00}&u_{01}v_{01}&u_{02}v_{00}&u_{02}v_{01}\\
    u_{00}v_{10}&u_{00}v_{11}&u_{01}v_{10}&u_{01}v_{11}&u_{02}v_{10}&u_{02}v_{11}\\
    \hline
    u_{10}v_{00}&u_{10}v_{01}&u_{11}v_{00}&u_{11}v_{01}&u_{12}v_{00}&u_{12}v_{01}\\
    u_{10}v_{10}&u_{10}v_{11}&u_{11}v_{10}&u_{11}v_{11}&u_{12}v_{10}&u_{12}v_{11}\\
    \hline
    u_{20}v_{00}&u_{20}v_{01}&u_{21}v_{00}&u_{21}v_{01}&u_{22}v_{00}&u_{22}v_{01}\\
    u_{20}v_{10}&u_{20}v_{11}&u_{21}v_{10}&u_{21}v_{11}&u_{22}v_{10}&u_{22}v_{11}\\
     \end{array} \right). 
\end{align}
It is possible to invert the tensor operation by using the property that both $\mathbf{U}$ and $\mathbf{V}$ are stochastic, and therefore sums over columns of $\mathbf{V}$ or $\mathbf{U}$ contained in $\interactionM$ are cancelled out and can be used to isolate the remaining variable. As an example one may obtain $U_{00}=A_{00}+A_{10}$ and also as $U_{00}=A_{10}+A_{11}$. The same is true for $\mathbf{V}$ as $V_{00}=A_{00}+A_{20}+A_{40}$, $V_{00}=A_{02}+A_{22}+A_{42}$ and $V_{00}=A_{04}+A_{24}+A_{44}$. Since these equations are redundant, one may pick one at random and discard the rest. The expression at Appendix Eq. \eqref{sm_eq:sub}  selects always the first one. This technique will always output a $\mathbf{U},\mathbf{V}$ pair even if the tensor factorization does not exist. Therefore one should always check whether $\interactionM= \mathbf{U} \otimes \mathbf{V}$ holds. For the purpose of finding subsystem interactions, we recursively apply this procedure on different combinations of interaction partitions until no more subsystem interactions can be found.

\subsection{Complexity}
In this section we prove that $r$ is the fraction of the area behind $k^c(l)$.
From the definition of $k(l)=\sum_{\lambda (\interaction^i)=l} a_i$ we could obtain the cumulative as $k^c(l)=\sum_{\lambda (\interaction^i) \le l} a_i$. Instead, we are going to express the $k$ as a function of a coninuous variable $\phi \in [0,\lambda_{\text{MC}}]$ as $k=\sum a_i \delta(\phi-\lambda(\interaction^i))$. Then the cumulative $k^c(\phi)$ is
\begin{align*}
k^c(\phi) &= \int^{\phi}_{0} k(\phi') d\phi' \\
&= \int^{\phi}_{0} \sum a_i \delta(\phi'-\lambda(\interaction^i))d\phi' \\
&= \sum a_i \int^{\phi}_{0} \delta(\phi'-\lambda(\interaction^i))d\phi' \\
&= \sum a_i \theta(\phi - \lambda(\interaction^i)).
\end{align*}
Now the area behind $k^c(\phi)$ is
\begin{align*}
&\int^{\lambda_{\text{MC}}}_{0} k^c(\phi) d\phi = \int^{\lambda_{\text{MC}}}_{0} \sum a_i \theta(\phi - \lambda(\interaction^i)) d\phi \\
&= \sum a_i  \int^{\lambda_{\text{MC}}}_{0} \theta(\phi - \lambda(\interaction^i)) d\phi \\
&= \sum a_i \int^{\lambda_{\text{MC}}-\lambda(\interaction^i)}_{-\lambda(\interaction^i)} \theta(\phi') d\phi' \\
&= \sum a_i (\lambda_{\text{MC}}-\lambda(\interaction^i)) \\
&= \lambda_{\text{MC}} - \sum a_i \lambda(\interaction^i).
\end{align*}
The total area is simply $\lambda_{\text{MC}}$, therefore the share is obtained by dividing the expression above with $\lambda_{\text{MC}}$, which yields exactly $r$.

\onecolumngrid

\clearpage

\centerline{\large \textbf{Supplemental Material}}

In this document we showcase our formalism in different practical applications.

\setcounter{section}{0}

\section{Experiments with synthetic datasets}
In this section we elaborate on the experiments with synthetic datasets.

The procedure to carry out the first experiment in the main text is the following:
\begin{itemize}
\itemsep0em 
\item We create a transition matrix $\mathbf{T}$ whose elements are selected at random from a uniform distribution in [0,1].
\item We normalize $\mathbf{T}$.
\item We apply the decomposition algorithm to matrix $\mathbf{T}$ and obtain the interaction ensemble, the interaction matrices $\mathbf{A}$ and the coefficients $a$.
\item We compute the reducibility $r$.

\item We generate sequences of length $L$ using $\mathbf{T}$:
    \begin{itemize}
    \item For processes with order $o$ the first $o-1$ elements of every sequence are drawn from a uniform distribution in $X$.
    \item The next element, is selected at random from $X$ according to the distribution given by the column of $\T$ corresponding to $\G^{o-1}(o-1)$. 
    \item The procedure is repeated, the $\tau$th element is selected at random from $X$ according to the distribution of the column of $\T$ corresponding to $\G^{o-1}(\tau-1)$.
    \end{itemize}

\item We measure the statistics of the sequences we have generated, and encode them in the estimator transition matrix $\hat{\T}$.
\item We apply the decomposition algorithm to the estimator matrix $\hat{\T}$ and obtain the estimated interaction ensemble, all the estimated interaction matrices $\hat{\mathbf{A}}$ and the estimated coefficients $\hat{a}$.
\item We compute the estimated reducibility $\hat{r}$.

\item We compute the total variation distance between the original transition matrix $\mathbf{T}$ and the estimated one $\hat{\mathbf{T}}$.
\item We compute the total variation distance between the original probability mass function of the coefficients $a$ and the probability mass function of the estimated coefficients $\hat{a}$.
\item We compute the absolute value of the difference between the original reducibility $r$ and the estimated one $\hat{r}$.
\end{itemize}

The length of the sequence is given by $L=\eta \prod_{n \in \mathcal{N}}|X_n|^{\order}$, where $\eta$ is a scale factor that determines the average frequency of each unique string for a uniform model. In other words, $\eta$ is the expected average number of repetitions of every element of $\times_{n \in \mathcal{N}}X_n^\order$ in the sequence. Therefore $\eta$ gives an idea of the available amount of data. We use it to tune the $\sigma(\T,\hat{\T})$ distance.\\

We also show two additional experiments to validate the robustness of our algorithm to extract interactions from data. First we replicate the experiment in the main text with a different order of $o=3$, see Supplementary Fig. \ref{sup_fig:val0}. The results are aligned with those in the main text, and show that one can obtain a reducible ensemble of interactions with a good predicting capacity even when there is not enough data to accurately infer the ground truth interactions.\\

\begin{figure}[h]
\centering
\subfloat[]{\includegraphics[width=0.25\textwidth]{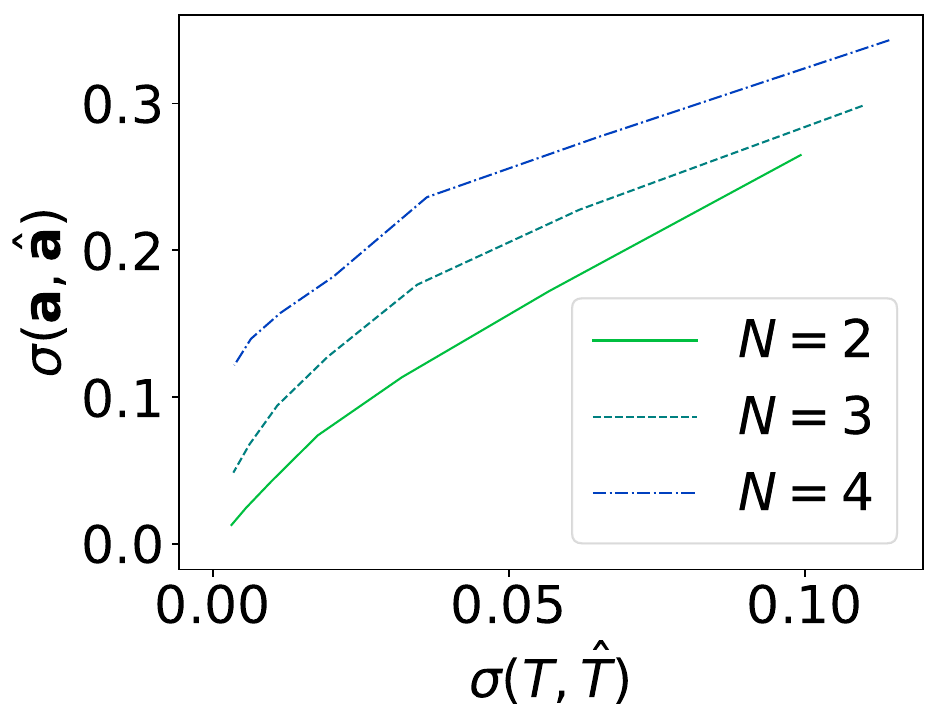}}
\subfloat[]{\includegraphics[width=0.25\textwidth]{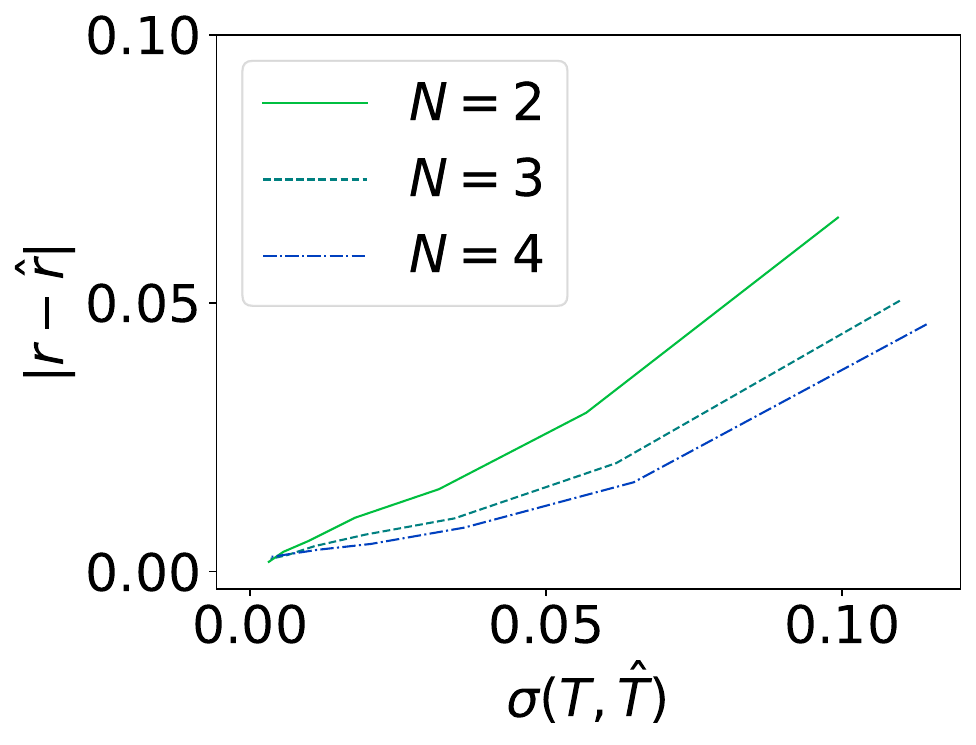}}\\[-0.25cm]
\caption{We produce synthetic sequences and compute the statistical distance between the original transition matrices $\T$, (a) interactions $\mathbf{a}$ and (b) reducibilities $r$, and the ones inferred from the data $\hat{\T}, \hat{\mathbf{a}}, \hat{r}$ for a fixed alphabet $|X_n|=2$ and order $o=3$. Results reported here correspond to averages over $100$ realizations.} 
\label{sup_fig:val0}
\end{figure}

We have carried out a second experiment to explore alphabets with $|X_n| \neq 2$. The procedure is the same as the one explained in the main text: we first randomly generate a transition matrix $\T$ and extract the ground truth values for interaction coefficients $a$ and reducibilities $r$. We then create sequences of length $L=\eta \prod_{n \in \mathcal{N}}|X_n|^{\order}$ from $\T$, and use them to infer $\hat\T$. We apply our algorithm to $\hat\T$ and obtain $\hat{a}$ and $\hat{r}$. We run $100$ simulations for each different $(X,N,\order)$ tuple. See the results in Supplementary Fig. \ref{sup_fig:val}. 

\begin{figure}[h]
\centering
\subfloat[]{\includegraphics[width=0.25\textwidth]{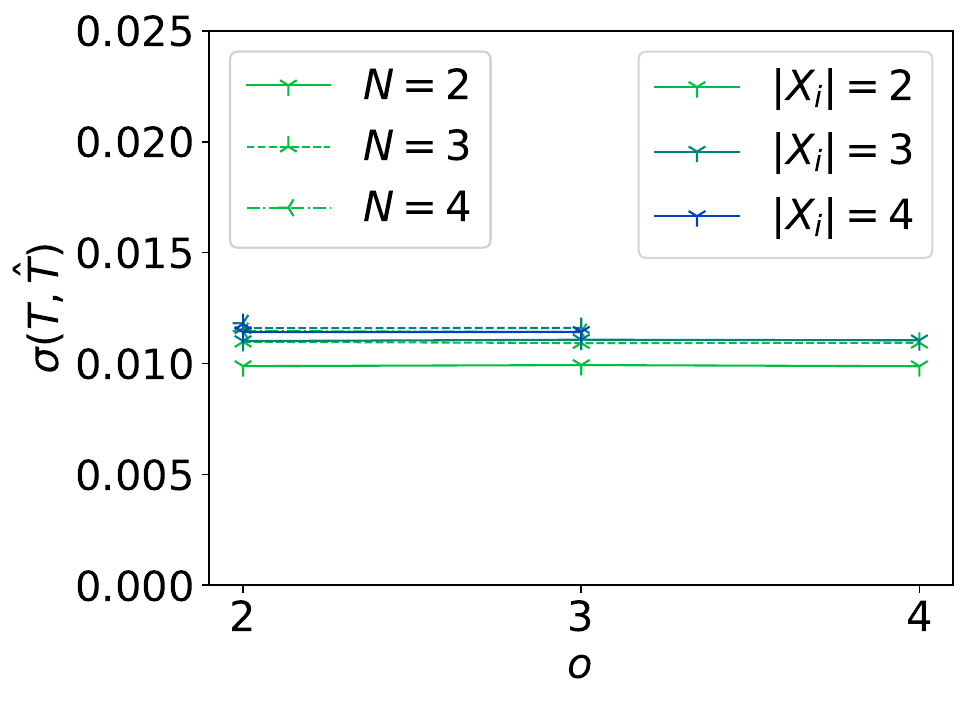}}
\subfloat[]{\includegraphics[width=0.25\textwidth]{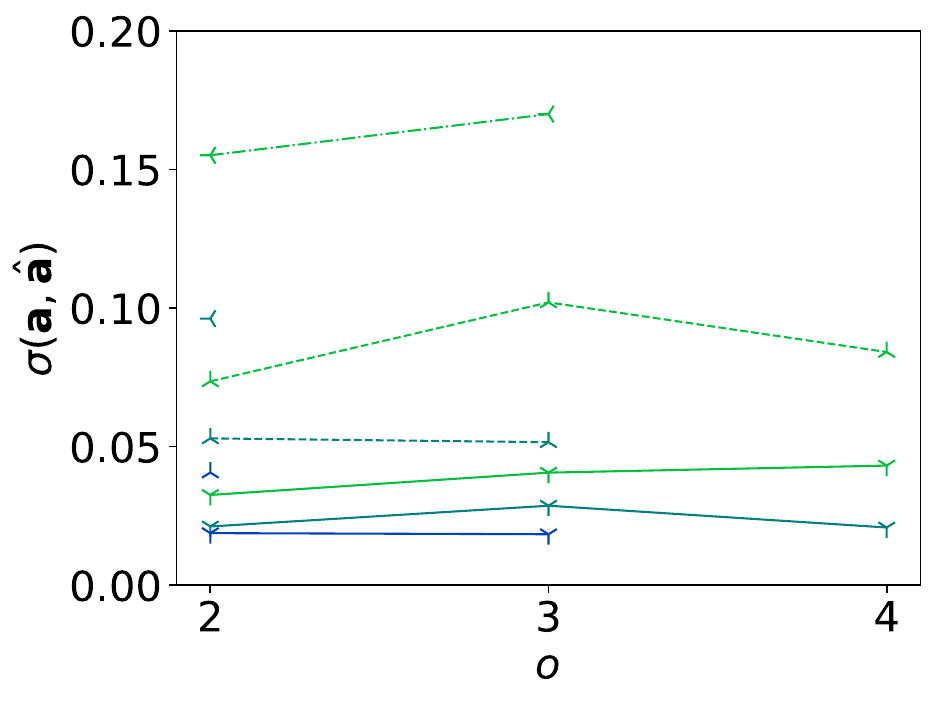}}
\subfloat[]{\includegraphics[width=0.25\textwidth]{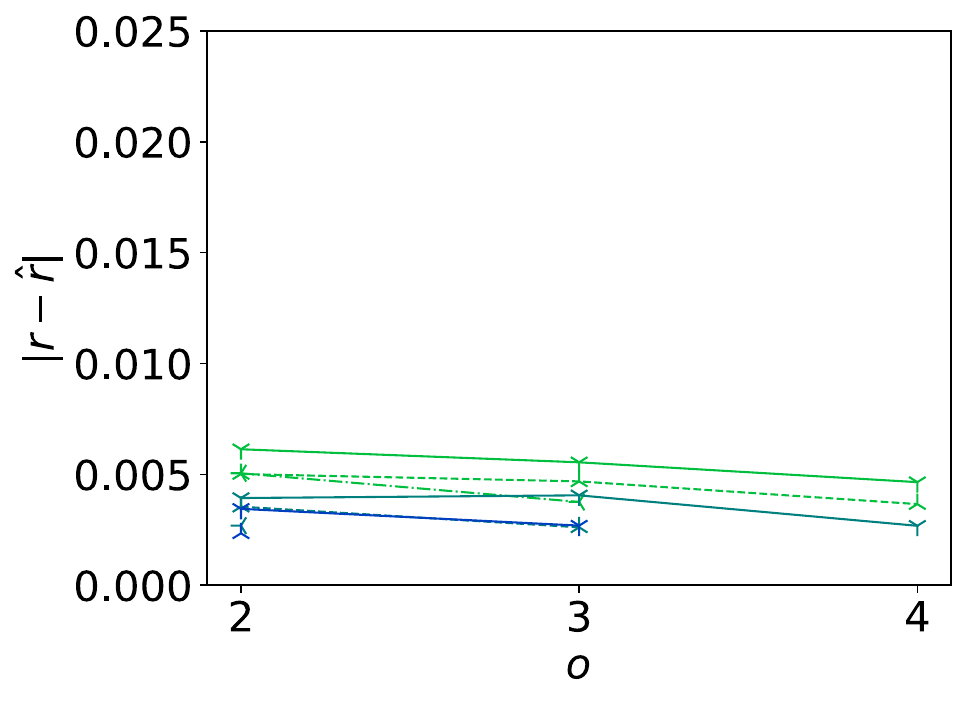}}\\[-0.25cm]
\caption{We produce synthetic sequences with length $L=\eta \prod_{n \in \mathcal{N}}|X_n|^{\order}$ and compute the total variation distance $\sigma$ between the original (a) transition matrices (b) interactions (c) reducibilities, and the ones inferred from the data for a fixed $\eta=1000$. Results reported here correspond to averages over $100$ realizations for each $(X,N,\order)$ tuple. The curve legends depicted in (a) also apply to the plots in (b) and (c).} 
\label{sup_fig:val}
\end{figure}

We observe again the trend explained in the main text: despite interactions being very susceptible to statistical errors, the reducibility is not. This implies that when studying the dynamics of a time-ordered and multibody system, the reducibility of inferred interaction decomposition will be close to the correct value even if interaction themselves are different.

\section{Experiments with real datasets}
\subsection{Glauber dynamics}
We shall provide a toy model for the analysis of Glauber dynamics \cite{glauber} using our formalism. Consider a system of $N=3$ nodes with and alphabet $X_{n}=\{-1,1\}$. Let us assume nodes are arranged in a triangle such that all nodes influence each other. The Glauber dynamics is asynchronous, i.e., acts on a single node at a given time step, and updates its state depending on the neighbouring nodes. More precisely, a node $n$ will change its state, $s_n(t)=-s_n(t-1)$ with a probability $\pi$ that depends on the sum of the neighbour node states $\Phi=\sum_{m \in \mathcal{N} | m \neq n} s_m(t-1)$ and the inverse of the temperature $z$,
\begin{equation}
\pi[s_n(t)=-s_n(t-1)]=\frac{e^{-2 z s_n(t-1) \Phi }}{1+e^{- 2 z s_n(t-1) \Phi }}.
\label{sup_eq:dyn}
\end{equation}

Given the asynchronous character of the dynamics, we compute the joint evolution of the system by adding the contributions of each node with a uniform probability. Therefore, the joint transition matrix $\T$ is the sum of individual transition matrices $\T_n$. We select a maximum $\order=2$, as Glauber dynamics requires no memory. Accordingly, the transition matrix is simply an input-output $8 \times 8$ matrix, where the columns account for all the possible $s(t-1)$ and the rows account for $s(t)$, such that matrix elements correspond to $\pi(s(t)|s(t-1))$. For a better readability let us define the function $\alpha(z)$ as $\alpha \equiv \frac{e^{-4 z}}{1+e^{-4 z}}$. 

\begin{table}[h] 
\renewcommand{\arraystretch}{1.2}
\begin{tabular}{c|c|c|c}
\centering
 & $n=0$ & $n=1$ & $n=2$ \\ 
\hline
$s(t-1)$ & $s(t)$ & $s(t)$ & $s(t)$  \\
\hline
$(-1,-1,-1)$ & $(1-\alpha)(-1,-1,-1)+\alpha(1,-1,-1)$ & $(1-\alpha)(-1,-1,-1)+\alpha(-1,1,-1)$ & $(1-\alpha)(-1,-1,-1)+\alpha(-1,-1,1)$ \\
$(-1,-1,1)$ & $0.5(-1,-1,1)+0.5(1,-1,1)$ & $0.5(-1,-1,1)+0.5(-1,1,1)$ & $\alpha(-1,-1,1)+(1-\alpha)(-1,-1,-1)$ \\
$(-1,1,-1)$ & $0.5(-1,1,-1)+0.5(1,1,-1)$ & $\alpha(-1,1,-1)+(1-\alpha)(-1,-1,-1)$ & $0.5(-1,1,-1)+0.5(-1,1,1)$ \\
$(-1,1,1)$ & $\alpha(-1,1,1)+(1-\alpha)(1,1,1)$ & $0.5(-1,1,1)+0.5(-1,-1,1)$ & $0.5(-1,1,1)+0.5(-1,1,-1)$ \\
$(1,-1,-1)$ & $\alpha(1,-1,-1)+(1-\alpha)(-1,-1,-1)$ & $0.5(1,-1,-1)+0.5(1,1,-1)$ & $0.5(1,-1,-1)+0.5(1,-1,1)$ \\
$(1,-1,1)$ & $0.5(1,-1,1)+0.5(-1,-1,1)$ & $\alpha(1,-1,1)+(1-\alpha)(1,1,1)$ & $0.5(1,-1,1)+0.5(1,-1,-1)$ \\
$(1,1,-1)$ & $0.5(1,1,-1)+0.5(-1,1,-1)$ & $0.5(1,1,-1)+0.5(1,-1,-1)$ & $\alpha(1,1,-1)+(1-\alpha)(1,1,1)$ \\
$(1,1,1)$ & $(1-\alpha)(1,1,1)+\alpha(-1,1,1)$ & $(1-\alpha)(1,1,1)+\alpha(1,-1,1)$ & $(1-\alpha)(1,1,1)+\alpha(1,1,-1)$ \\
\hline
\end{tabular}
\caption{Transitions within the Glauber dynamics for three nodes.}
\label{tab:glauber}
\end{table}

We now explicitly compute all transition probabilities for each node at Table \ref{tab:glauber}, and we use these to construct transition matrices. We denote by $\T_0$, $\T_1$ and $\T_2$ the transition matrix corresponding to each node, which read
\begin{eqnarray*}
\tiny
\left( \begin{array}{cccccccc}
                            1-\alpha & 0 & 0 & 0 & 1-\alpha & 0 & 0 & 0\\
                            0 & 1/2 & 0 & 0 & 0 & 1/2 & 0 & 0\\
                            0 & 0 & 1/2 & 0 & 0 & 0 & 1/2 & 0\\
                            0 & 0 & 0 & \alpha & 0 & 0 & 0 & \alpha\\
                            \alpha & 0 & 0 & 0 & \alpha & 0 & 0 & 0\\
                            0 & 1/2 & 0 & 0 & 0 & 1/2 & 0 & 0\\
                            0 & 0 & 1/2 & 0 & 0 & 0 & 1/2 & 0\\
                            0 & 0 & 0 & 1-\alpha & 0 & 0 & 0 & 1-\alpha\\
\end{array} \right),    
\left(  \begin{array}{cccccccc} 
                            1-\alpha & 0 & 1-\alpha & 0 & 0 & 0 & 0 & 0\\
                            0 & 1/2 & 0 & 1/2 & 0 & 0 & 0 & 0\\
                            \alpha & 0 & \alpha & 0 & 0 & 0 & 0 & 0\\
                            0 & 1/2 & 0 & 1/2 & 0 & 0 & 0 & 0\\
                            0 & 0 & 0 & 0 & 1/2 & 0 & 1/2 & 0\\
                            0 & 0 & 0 & 0 & 0 & \alpha & 0 & \alpha\\
                            0 & 0 & 0 & 0 & 1/2 & 0 & 1/2 & 0\\
                            0 & 0 & 0 & 0 & 0 & 1-\alpha & 0 & 1-\alpha\\    
\end{array} \right),
\left( \begin{array}{cccccccc}
                            1-\alpha & 1-\alpha & 0 & 0 & 0 & 0 & 0 & 0\\
                            \alpha & \alpha & 0 & 0 & 0 & 0 & 0 & 0\\
                            0 & 0 & 1/2 & 1/2 & 0 & 0 & 0 & 0\\
                            0 & 0 & 1/2 & 1/2 & 0 & 0 & 0 & 0\\
                            0 & 0 & 0 & 0 & 1/2 & 1/2 & 0 & 0\\
                            0 & 0 & 0 & 0 & 1/2 & 1/2 & 0 & 0\\
                            0 & 0 & 0 & 0 & 0 & 0 & \alpha & \alpha\\
                            0 & 0 & 0 & 0 & 0 & 0 & 1-\alpha & 1-\alpha\\
\end{array} \right).
\end{eqnarray*}
The joint transition matrix is obtained with $\T=( \T_0 + \T_1 + \T_2 )/3$
\begin{equation*}
\tiny
\T=\left( \begin{array}{cccccccc}
    1-\alpha & (1-\alpha)/3 & (1-\alpha)/3 & 0 & (1-\alpha)/3 & 0 & 0 & 0\\
    \alpha/3 & (1+\alpha)/3 & 0 & 1/6 & 0 & 1/6 & 0 & 0\\
    \alpha/3 & 0 & (1+\alpha)/3 & 1/6 & 0 & 0 & 1/6 & 0\\
    0 & 1/6 & 1/6 & (1+\alpha)/3 & 0 & 0 & 0 & \alpha/3\\
    \alpha/3 & 0 & 0 & 0 & (1+\alpha)/3 & 1/6 & 1/6 & 0\\
    0 & 1/6 & 0 & 0 & 1/6 & (1+\alpha)/3 & 0 & \alpha/3\\
    0 & 0 & 1/6 & 0 & 1/6 & 0 & (1+\alpha)/3 & \alpha/3\\
    0 & 0 & 0 & (1-\alpha)/3 & 0 & (1-\alpha)/3 & (1-\alpha)/3 & 1-\alpha\\
\end{array} \right).
\end{equation*}

Now that we have computed $\T$ we will apply our decomposition algorithm explained in paragraph $c$ to obtain an equivalent interaction ensemble. Thorough this analysis we will use the notation $\interactionM^{\q}$ and $\R^{\q}$ to denote an interaction matrix of type $\q$ and the auxiliary matrix of type $\q$. Notice that the dimensions of the system we are studying coincide with those at Fig. $2$ in the main manuscript. Thus Fig. $2$ in the main text displays the interactions to be checked at each interaction order. We are also going to use the following notation:
\begin{equation*}
\{\T_{[i_{\min,},i_{\max}][j_{\min},j_{\max}]}\}=\{T_{ij}|i \in \{i_{\min},...,i_{\max}\}, j \in \{j_{\min},...,j_{\max}\}\},
\end{equation*}
to express a set of matrix elements in a compact form. Let us derive the decomposition:

\begin{itemize}
\itemsep0em 
\item $\omega=0$: The interaction with type $\q=(0,0,0)$ is the only one at this order. The auxiliary matrix $\R^{(000)}$ is simply the minimal element in $\T$, so $\R^{(000)}=(0)$. This means we cannot extract interactions at $\omega=0$.
\item $\omega=1$: We have three interaction types, $\q=(1,0,0)$, $\q=(0,1,0)$ and $\q=(0,0,1)$. In the three cases we have a matrix with a single column and two rows, whose elements are computed following the algorithm explained in the Appendix of the main text.
\begin{equation*}
\R^{(100)}=\left( \begin{array}{c} \min \{\T_{[0,3][0,7]}\} \\ \min\{\T_{[4,7][0,7]}\} \end{array} \right) = \left( \begin{array}{c} 0 \\ 0 \end{array} \right).
\end{equation*}
The same result is found for the remaining two types, and thus we cannot extract any interaction at this order.
\item $\omega=2$: We have six possible interaction types, $\q=(1,1,0)$, $\q=(1,0,1)$, $\q=(0,1,1)$, $\q=(2,0,0)$, $\q=(0,2,0)$ and $\q=(0,0,2)$. Again, all we have to do is to apply the algorithm explained in the Appendix of the main text to detect which matrix elements of $\T$ correspond to those in $\R$, 
\begin{equation*}
\R^{(110)}=\left( \begin{array}{c} \min\{\T_{[0,1][0,7]}\} \\
                                   \min\{\T_{[2,3][0,7]}\} \\
                                   \min\{\T_{[4,5][0,7]}\} \\
                                   \min\{\T_{[6,7][0,7]}\}
\end{array} \right) = \left( \begin{array}{c} 0 \\ 0 \\ 0 \\ 0 \end{array} \right).
\end{equation*}
The same result is achieved for $\R^{(101)}$ and $\R^{(011)}$, so the dynamics does not contain those interactions either. Let us go for $\R^{(200)}$:
\begin{equation*}
\R^{(200)}= \left( \begin{array}{cc} \min\{\T_{[0,3][0,3]}\} & \min\{\T_{[0,3][4,7]}\} \\
                                     \min\{\T_{[4,7][0,3]}\} & \min\{\T_{[4,7][4,7]}\}
\end{array} \right) = \left( \begin{array}{cc} 0 & 0 \\ 0 & 0 \end{array} \right).
\end{equation*}
$\R^{(020)}$ and $\R^{(002)}$ yield the same results. Thus, we have no interactions and we should go for the next order.
\item $\omega=3$: We have seven interaction types at this order $\q=(1,1,1)$, $\q=(2,1,0)$, $\q=(2,0,1)$, $\q=(0,2,1)$, $\q=(1,2,0)$, $\q=(0,1,2)$ and $\q=(1,0,2)$. Let us obtain the auxiliary matrices for $\q=(1,1,1)$ and $\q=(2,1,0)$:
\begin{equation*}
\R^{(111)} = \left( \begin{array}{c} \min\{\T_{[0][0,7]}\} \\ \min\{\T_{[1][0,7]}\} \\ \min\{\T_{[2][0,7]}\} \\ \min\{\T_{[3][0,7]}\} \\
                                     \min\{\T_{[4][0,7]}\} \\ \min\{\T_{[5][0,7]}\} \\ \min\{\T_{[6][0,7]}\} \\ \min\{\T_{[7][0,7]}\} \end{array} \right)= \left( \begin{array}{c} 0\\0\\0\\0\\0\\0\\0\\0 \end{array} \right),
\hspace{0.5cm}
\R^{(210)} = \left( \begin{array}{cc} \min\{\T_{[0,1][0,3]}\} & \min\{\T_{[0,1][4,7]}\} \\
                                      \min\{\T_{[2,3][0,3]}\} & \min\{\T_{[2,3][4,7]}\} \\
                                      \min\{\T_{[4,5][0,3]}\} & \min\{\T_{[4,5][4,7]}\} \\
                                      \min\{\T_{[6,7][0,3]}\} & \min\{\T_{[6,7][4,7]}\}
\end{array} \right)= \left( \begin{array}{cc} 0&0\\0&0\\0&0\\0&0 \end{array} \right).
\end{equation*}
None of these yields any interaction. The remaining ones at order $\omega=3$ have the same matrix dimensions as $\q~=~(2,1,0)$, and are computed similarly. They are all null.
\item $\omega=4$: We have six interaction types divided into two families $\q=(2,1,1)$, $\q=(1,2,1)$, $\q=(1,1,2)$ and $\q=(2,2,0)$, $\q=(2,0,2)$, $\q=(0,2,2)$. We now show the calculation for an interaction of each family:
\begin{equation*}
\R^{(211)}=\left( \begin{array}{cc} \min\{\T_{[0][0,3]}\} & \min\{\T_{[0][4,7]}\} \\
                                      \min\{\T_{[1][0,3]}\} & \min\{\T_{[1][4,7]}\} \\
                                      \min\{\T_{[2][0,3]}\} & \min\{\T_{[2][4,7]}\} \\
                                      \min\{\T_{[3][0,3]}\} & \min\{\T_{[3][4,7]}\} \\
                                      \min\{\T_{[4][0,3]}\} & \min\{\T_{[4][4,7]}\} \\
                                      \min\{\T_{[5][0,3]}\} & \min\{\T_{[5][4,7]}\} \\
                                      \min\{\T_{[6][0,3]}\} & \min\{\T_{[6][4,7]}\} \\
                                      \min\{\T_{[7][0,3]}\} & \min\{\T_{[7][4,7]}\} \end{array} \right) = \left( \begin{array}{cc} 0&0\\0&0\\0&0\\0&0\\0&0\\0&0\\0&0\\0&0 \end{array} \right).
\end{equation*}  
The remaining two auxiliary matrices of this same family, $\R^{(121)}$ and $\R^{(112)}$, are also incompatible with these statistics:
\begin{eqnarray*}
\tiny
\R^{(220)}&=&\left( \begin{array}{cccc} \min\{\T_{[0,1][0,1]}\} & \min\{\T_{[0,1][2,3]}\} & \min\{\T_{[0,1][4,5]}\} & \min\{\T_{[0,1][6,7]}\} \\
                                      \min\{\T_{[2,3][0,1]}\} & \min\{\T_{[2,3][2,3]}\} & \min\{\T_{[2,3][4,5]}\} & \min\{\T_{[2,3][6,7]}\} \\
                                      \min\{\T_{[4,5][0,1]}\} & \min\{\T_{[4,5][2,3]}\} & \min\{\T_{[4,5][4,5]}\} & \min\{\T_{[4,5][6,7]}\} \\
                                      \min\{\T_{[6,7][0,1]}\} & \min\{\T_{[6,7][2,3]}\} & \min\{\T_{[6,7][4,5]}\} & \min\{\T_{[6,7][6,7]}\} \end{array} \right).
\end{eqnarray*}
For the first time we find a non-zero auxiliary matrix with the following structure:
\begin{eqnarray*}
\R^{(220)}=\left( \begin{array}{cccc} R_{00}&0&0&0\\0&R_{11}&0&0\\0&0&R_{22}&0\\0&0&0&R_{33} \end{array} \right), \hspace{0.5cm} 
\begin{array}{c} R_{00}=R_{33}=\min\{1-\alpha,(1-\alpha)/3,(1+\alpha)/3,\alpha/3\}, \\ R_{11}=R_{22}=\min\{(1+\alpha)/3,1/6\}. \end{array}
\end{eqnarray*}
One can show that the auxiliary matrices of the remaining two interactions, $\R^{(202)}$ and $\R^{(022)}$ are identical to $\R^{(220)}$. Interaction matrices $\interactionM^{(220)}$, $\interactionM^{(202)}$ and $\interactionM^{(022)}$ are obtained by normalizing the corresponding auxiliary matrices
\begin{equation}
\interactionM^{(220)}=\interactionM^{(202)}=\interactionM^{(022)}=\left( \begin{array}{cccc} 1&0&0&0\\0&1&0&0\\0&0&1&0\\0&0&0&1 \end{array} \right).
\label{eq:random}
\end{equation} 
Notice that although interactions matrices are the same for these three interactions, when we nest them in an interaction type $\q=(2,2,2)$, the one corresponding to $\T$, they are different:
\begin{equation*}
\tiny
\bar{\interactionM}^{(022)}=\frac{1}{2} \left( \begin{array}{cccccccc} 1&0&0&0&1&0&0&0\\0&1&0&0&0&1&0&0\\0&0&1&0&0&0&1&0\\0&0&0&1&0&0&0&1\\1&0&0&0&1&0&0&0\\0&1&0&0&0&1&0&0\\0&0&1&0&0&0&1&0\\0&0&0&1&0&0&0&1 \end{array} \right),
\hspace{0.5cm}
\bar{\interactionM}^{(202)}=\frac{1}{2} \left( \begin{array}{cccccccc} 1&0&1&0&0&0&0&0\\0&1&0&1&0&0&0&0\\1&0&1&0&0&0&0&0\\0&1&0&1&0&0&0&0\\0&0&0&0&1&0&1&0\\0&0&0&0&0&1&0&1\\0&0&0&0&1&0&1&0\\0&0&0&0&0&1&0&1 \end{array} \right),
\hspace{0.5cm}
\bar{\interactionM}^{(220)}=\frac{1}{2} \left( \begin{array}{cccccccc} 1&1&0&0&0&0&0&0\\1&1&0&0&0&0&0&0\\0&0&1&1&0&0&0&0\\0&0&1&1&0&0&0&0\\0&0&0&0&1&1&0&0\\0&0&0&0&1&1&0&0\\0&0&0&0&0&0&1&1\\0&0&0&0&0&0&1&1 \end{array} \right).
\end{equation*} 

Interaction coefficients $a^{(220)}$, $a^{(202)}$ and $a^{(022)}$ are extracted from the non-zero elements in the auxiliary matrices. In this particular case we will extract three interactions in a single operation and with the same coefficient. Thus, we are looking for the highest possible coefficient $a^{(220)}=a^{(202)}=a^{(022)}=\beta$, such that $\T-\beta(\bar{\interactionM}^{(220)}+\bar{\interactionM}^{(202)}+\bar{\interactionM}^{(022)})$ doesn't have negative elements. We find $\beta=2\alpha/3$ to be the best option, because it creates zeros in all those elements of $\T$ associated with $\alpha/3$. This is derived from $R_{00},R_{33}$ and $R_{11},R_{22}$. Before proceeding with the decomposition we shall remove from $\T$ the interactions that have just found.
\begin{equation*}
\tiny
\tilde{\T}=T- \frac{2 \alpha}{3} (\bar{\interactionM}^{(220)}+\bar{\interactionM}^{(202)}+\bar{\interactionM}^{(022)}) = (1-2\alpha)
\left( \begin{array}{cccccccc}
       1&1/3&1/3&0&1/3&0&0&0\\
       0&1/3&0&1/6&0&1/6&0&0\\
       0&0&1/3&1/6&0&0&1/6&0\\
       0&1/6&1/6&1/3&0&0&0&0\\
       0&0&0&0&1/3&1/6&1/6&0\\
       0&1/6&0&0&1/6&1/3&0&0\\
       0&0&1/6&0&1/6&0&1/3&0\\
       0&0&0&1/3&0&1/3&1/3&1\\
       \end{array}\right).
\end{equation*}
Now we continue the decomposition of $\tilde{\T}$ with the next interaction order.
\item $\omega=5$: There are three possible interaction types at this order, $\q=(2,2,1)$, $\q=(2,1,2)$ and $\q=(1,2,2)$. Following the same procedure explained above we have
\begin{equation*}
\R^{(221)}= \left( \begin{array}{cccc} 
    \min{\tilde{\T}_{[0][0,1]}}&\min{\tilde{\T}_{[0][2,3]}}&\min{\tilde{\T}_{[0][4,5]}}&\min{\tilde{\T}_{[0][6,7]}} \\
    \min{\tilde{\T}_{[1][0,1]}}&\min{\tilde{\T}_{[1][2,3]}}&\min{\tilde{\T}_{[1][4,5]}}&\min{\tilde{\T}_{[1][6,7]}} \\
    \min{\tilde{\T}_{[2][0,1]}}&\min{\tilde{\T}_{[2][2,3]}}&\min{\tilde{\T}_{[2][4,5]}}&\min{\tilde{\T}_{[2][6,7]}} \\
    \min{\tilde{\T}_{[3][0,1]}}&\min{\tilde{\T}_{[3][2,3]}}&\min{\tilde{\T}_{[3][4,5]}}&\min{\tilde{\T}_{[3][6,7]}} \\
    \min{\tilde{\T}_{[4][0,1]}}&\min{\tilde{\T}_{[4][2,3]}}&\min{\tilde{\T}_{[4][4,5]}}&\min{\tilde{\T}_{[4][6,7]}} \\
    \min{\tilde{\T}_{[5][0,1]}}&\min{\tilde{\T}_{[5][2,3]}}&\min{\tilde{\T}_{[5][4,5]}}&\min{\tilde{\T}_{[5][6,7]}} \\
    \min{\tilde{\T}_{[6][0,1]}}&\min{\tilde{\T}_{[6][2,3]}}&\min{\tilde{\T}_{[6][4,5]}}&\min{\tilde{\T}_{[6][6,7]}} \\
    \min{\tilde{\T}_{[7][0,1]}}&\min{\tilde{\T}_{[7][2,3]}}&\min{\tilde{\T}_{[7][4,5]}}&\min{\tilde{\T}_{[7][6,7]}} \\
\end{array} \right),
\end{equation*} 
which results in a matrix with some non-zero elements
\begin{equation*}
\R^{(221)}=\left( \begin{array}{cccc} R_{00}&0&0&0 \\ 0&0&0&0 \\ 0&R_{21}&0&0 \\ 0&R_{31}&0&0 \\ 0&0&R_{42}&0 \\ 0&0&R_{52}&0 \\ 0&0&0&0 \\ 0&0&0&R_{73} \end{array} \right),
\hspace{1cm} \begin{array}{c} R_{00}=R_{73}=(1-2\alpha) \min\{1,1/3\}, \\ R_{21}=R_{31}=R_{42}=R_{52}=(1-2\alpha) \min\{1/3,1/6\}. \end{array}
\end{equation*} 
$\R^{(212)}$ and $\R^{(122)}$ contain the same non-zero elements, although they are located in different positions. We compute the interaction matrices by normalizing the auxiliary matrices:
\begin{equation}
\tiny
\interactionM^{(221)}= \left( \begin{array}{cccc} 1&0&0&0 \\ 0&0&0&0 \\ 0&1/2&0&0 \\ 0&1/2&0&0 \\ 0&0&1/2&0 \\ 0&0&1/2&0 \\ 0&0&0&0 \\ 0&0&0&1 \end{array} \right),
\interactionM^{(212)}= \left( \begin{array}{cccc} 1&0&0&0 \\ 0&1/2&0&0 \\ 0&0&0&0 \\ 0&1/2&0&0 \\ 0&0&1/2&0 \\ 0&0&0&0 \\ 0&0&1/2&0 \\ 0&0&0&1 \end{array} \right),
\interactionM^{(122)}= \left( \begin{array}{cccc} 1&0&0&0 \\ 0&1/2&0&0 \\ 0&0&1/2&0 \\ 0&0&0&0 \\ 0&0&0&0 \\ 0&1/2&0&0 \\ 0&0&1/2&0 \\ 0&0&0&1 \end{array} \right),
\label{eq:magnet}
\end{equation} 
which we then nest into type $\q=(2,2,2)$:
\begin{equation*}
\tiny
\bar{\interactionM}^{(221)}= \left( \begin{array}{cccccccc} 1&1&0&0&0&0&0&0 \\ 0&0&0&0&0&0&0&0 \\ 0&0&1/2&1/2&0&0&0&0 \\ 0&0&1/2&1/2&0&0&0&0 \\ 0&0&0&0&1/2&1/2&0&0 \\ 0&0&0&0&1/2&1/2&0&0 \\ 0&0&0&0&0&0&0&0 \\ 0&0&0&0&0&0&1&1 \end{array} \right),
\bar{\interactionM}^{(212)}= \left( \begin{array}{cccccccc} 1&0&1&0&0&0&0&0 \\ 0&1/2&0&1/2&0&0&0&0 \\ 0&0&0&0&0&0&0&0 \\ 0&1/2&0&1/2&0&0&0&0 \\ 0&0&0&0&1/2&0&1/2&0 \\ 0&0&0&0&0&0&0&0 \\ 0&0&0&0&1/2&0&1/2&0 \\ 0&0&0&0&0&1&0&1 \end{array} \right),
\bar{\interactionM}^{(122)}= \left( \begin{array}{cccccccc} 1&0&0&0&1&0&0&0 \\ 0&1/2&0&0&0&1/2&0&0 \\ 0&0&1/2&0&0&0&1/2&0 \\ 0&0&0&0&0&0&0&0 \\ 0&0&0&0&0&0&0&0 \\ 0&1/2&0&0&0&1/2&0&0 \\ 0&0&1/2&0&0&0&1/2&0 \\ 0&0&0&1&0&0&0&1 \end{array} \right).
\end{equation*} 
We want to add these three interactions to the ensemble with the same coefficient. Notice that this procedure is different from the one in the original decomposition algorithm, where interactions are extracted one by one, and therefore the information in the auxiliary matrices is enough to obtain the interaction coefficients. We are then looking for the maximal $\beta$ such that $\tilde{\T}-\beta(\interactionM^{(221)}+\interactionM^{(212)}+\interactionM^{(122)})$ does not contain negative elements. We find that $\beta=(1-2\alpha)/3$ is the best option. Indeed, once these three interactions are removed from $\tilde{T}$, one is left with a null matrix, and thus the decomposition process is finished.
\end{itemize}
Finally, the transition matrix of the system may be expressed as follows:
\begin{equation}
\T = \frac{2\alpha}{3} (\bar{\interactionM}^{(220)}+\bar{\interactionM}^{(202)}+\bar{\interactionM}^{(022)}) + \frac{(1-2\alpha)}{3} (\bar{\interactionM}^{(221)}+\bar{\interactionM}^{(212)}+\bar{\interactionM}^{(122)}).
\end{equation}
This is providing us useful information about the system. Firstly, we are not using any interaction of the maximal type $\q=(2,2,2)$. In physical terms, this means that it is not necessary to know the state of all nodes in order to predict the next time step of the dynamics. Secondly, the dynamics is shown to be the outcome of two opposite contributions. Let us study both those terms separately by using the information encoded in the associated transition matrices respectively in SM Eqs. \ref{eq:random} and \ref{eq:magnet}. The first term corresponds to a random uniform evolution of a node while the rest remain unchanged. This means, that if this interaction is activated the node adopt either $-1$ or $1$ with $1/2$ probability. The second term corresponds to the magnetization. Nodes will update their state to the one preferred by their neighbours if both neighbours have the same state. If they have opposite states the node will adopt either $-1$ or $1$ with $1/2$ probability. See SM Fig. \ref{sup_fig:glauber3} for a graphical representation.

\begin{figure}[htb]
\captionsetup[subfigure]{labelformat=empty}
\centering
\subfloat[]{\includegraphics[width=\textwidth]{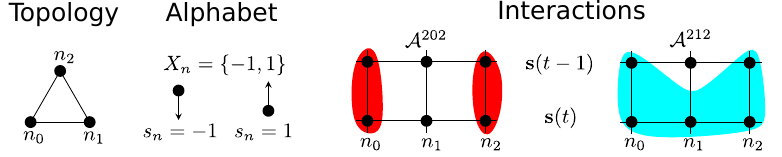}}
\caption{Interaction ensemble of Glauber dynamics. Let us assume we have a system of nodes coupled in a unidimensional ring. For simplicity let us start with $N=3$. Nodes are in a $-1$ or $+1$ state, corresponding to a classical and simplified representation of the spin of a particle. We compute the transition probabilities with SM Eq. \eqref{sup_eq:dyn} and decompose the dynamics into two families of interactions, where each family contains an interaction to update every node, while the rest of nodes remain in the same state. We illustrate $\interaction^{202}$ and $\interaction^{212}$, both responsible for updating $n_1$. In our figure rows account for the time, and columns account for the nodes. $\interaction^{202}$ tells us that $s_1(t)$ is determined randomly with uniform probability, as $q_1=0$, and therefore no information about the system is required to update $n_1$. $\interaction^{212}$ tells us that $\s(t)$ is a function of $s_0(t-1)$ and $s_2(t-1)$, but not a function of $s_1(t-1)$, and therefore the system is reducible.}  
\label{sup_fig:glauber3}
\end{figure}

Our conjecture for closed rings of $N$ nodes is
\begin{equation}
\T = \frac{2\alpha}{N} \left( \sum_{p \in \mathcal{P}[02...2]} \bar{\interactionM}^{p} \right) + \frac{(1-2\alpha)}{N} \left( \sum_{p \in \mathcal{P}[12...2]} \bar{\interactionM}^{p} \right),
\label{eq:conj}
\end{equation}
where $\mathcal{P[\q]}$ is the set with all permutations without repetitions of the components of $\q$. This expression is deduced from the transition probabilities for the $N$ node ring at Table \ref{tab:glauberN}. These show that individual transition matrices $T_n$ are essentially equivalent to those studied for $N=3$, as nodes that are not in the neighbourhood of a given node $n$ are not affected nor affect the update. Therefore, one can decompose the individual transition matrix $T_n$ into the random and magnetization interactions for the nodes in the neighbourhood, and with identity matrices, $q_n=2$, for nodes not in the neighbourhood.\\  
 
\begin{table}[h] 
\renewcommand{\arraystretch}{1.2}
\begin{tabular}{c|c}
\centering 
$s(t-1)$ & $s(t)$  \\
\hline
$(s_0,s_1,...,-1,-1,-1,...,s_{N-2},s_{N-1})$ & $(1-\alpha)(s_0,s_1,...,-1,-1,-1,...,s_{N-2},s_{N-1})+\alpha(s_0,s_1,...,-1,1,-1,...,s_{N-2},s_{N-1})$ \\
$(s_0,s_1,...,-1,-1,1,...,s_{N-2},s_{N-1})$ & $0.5(s_0,s_1,...,-1,-1,1,...,s_{N-2},s_{N-1})+0.5(s_0,s_1,...,-1,1,1,...,s_{N-2},s_{N-1})$ \\
$(s_0,s_1,...,1,-1,-1,...,s_{N-2},s_{N-1})$ & $0.5(s_0,s_1,...,1,-1,-1,...,s_{N-2},s_{N-1})+0.5(s_0,s_1,...,1,1,-1,...,s_{N-2},s_{N-1})$ \\
$(s_0,s_1,...,1,-1,1,...,s_{N-2},s_{N-1})$ & $\alpha(s_0,s_1,...,1,-1,1,...,s_{N-2},s_{N-1})+(1-\alpha)(s_0,s_1,...,1,1,1,...,s_{N-2},s_{N-1})$ \\
$(s_0,s_1,...,-1,1,-1,...,s_{N-2},s_{N-1})$ & $\alpha(s_0,s_1,...,-1,1,-1,...,s_{N-2},s_{N-1})+(1-\alpha)(s_0,s_1,...,-1,-1,-1,...,s_{N-2},s_{N-1})$ \\
$(s_0,s_1,...,-1,1,1,...,s_{N-2},s_{N-1})$ & $0.5(s_0,s_1,...,-1,1,1,...,s_{N-2},s_{N-1})+0.5(s_0,s_1,...,-1,-1,1,...,s_{N-2},s_{N-1})$ \\
$(s_0,s_1,...,1,1,-1,...,s_{N-2},s_{N-1})$ & $0.5(s_0,s_1,...,1,1,-1,...,s_{N-2},s_{N-1})+0.5(s_0,s_1,...,1,-1,-1,...,s_{N-2},s_{N-1})$ \\
$(s_0,s_1,...,1,1,1,...,s_{N-2},s_{N-1})$ & $(1-\alpha)(s_0,s_1,...,1,1,1,...,s_{N-2},s_{N-1})+\alpha(s_0,s_1,...,1,-1,1,...,s_{N-2},s_{N-1})$ \\
\end{tabular}
\caption{Glauber dynamics on a $N$ node ring. Transitions for node $n_i$ as a function of nodes $n_{i-1}$ and $n_{i+1}$.}
\label{tab:glauberN}
\end{table}

This interpretation is compatible with the known phenomenology. In the two opposite extremes of high temperatures $\alpha \sim 1/2$ and low temperatures $\alpha \sim 0$ the system is either fully random or fully ordered. Moreover, we argue that it is only possible to reach a fully ordered system at zero temperature for the following reasons. Suppose we have a fully ordered system (all nodes have the same state) and the temperature is low but different from zero. Sooner or later a node will change its state. If the temperature is low, the probability of a second random event on the same node (activation of the first term) is small, therefore let us focus on the magnetization term. Once a disorder has been created there are two possibilities: the dynamics corrects the disorder or the dynamics propagates the disorder. For the correction to occur, the node whose state is opposite to that of their neighbours has to be updated. The dynamics is asynchronous, and nodes are updated with a uniform probability, thus we have $1/N$ for the correction to occur (neglecting the random term). However, if the neighbours are updated instead, $2/N$ probability, we have $1/2$ probability to propagate the error, see SM Eq. \eqref{eq:magnet}. Therefore, once a disorder has been created there is $1/N$ probability to correct it and $1/N$ probability to propagate it in a given time step. This scenario inhibits the emergence of ordered structures above zero temperature.

\begin{figure}[htb]
\captionsetup[subfigure]{labelformat=empty}
\centering
\subfloat[]{\includegraphics[width=\textwidth]{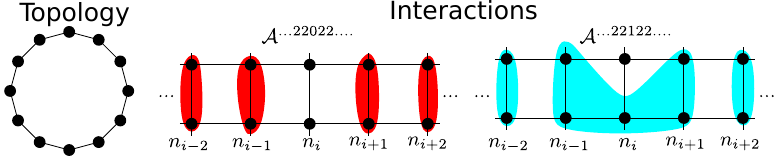}}
\caption{Glauber dynamics on a $N$ node ring. The dynamics is asynchronous, at each time step only a node is updated. The resulting interaction ensemble is fully symmetric, as all nodes have exactly the same two interactions associated. We shall call them random evolution and magnetization. They are the same to those in SM Fig. \ref{sup_fig:glauber3} with additional terms capturing the rest of nodes not in the neighbourhood. For nodes $n'$ with $|n'-n|>1$ the dynamics is an identity matrix: $q_n=2$ and an independent subsystem interaction, as in order to determine $s_n(t)$ one only needs to know $s_n(t-1)$.}  
\label{sup_fig:glauber8}
\end{figure}

\subsection{Traffic Jams}
We use our formalism to study transportation bottlenecks in the city of Sao Paulo between January 2010 and September 2018. The system is encoded in $N=5$ nodes with $|X_n|=2$ representing to the presence or absence of a traffic jam in the corresponding region of the city: $\{west,south,east,centre,north\}$. The dataset provides a record of traffic congestion events with a resolution of $30$ minutes, that we have transformed into a timeline of $L=153268$ states by using the alphabet $\{$ no jam $ \rightarrow 0$ , jam $\rightarrow 1\}$ for each node. We then create the joint state and apply our machinery to decompose the transition matrix in interactions. Based on $L=\eta \prod_{n \in \mathcal{N}}|X_n|^{\order}$, we set $\order^+=5$ as the maximum value of $o$. The AIC estimation yields a value of $\order=2$, for the multibody Markov chain. That implies that $\eta \sim 150$ which should yield values of $\sigma(\mathbf{a},\hat{\mathbf{a}})$ in the order of $0.3$ and values of $\sigma(\T,\hat{\T})$ in the order of $0.1$.\\ 

\begin{figure}[htb]
\captionsetup[subfigure]{labelformat=empty}
\centering
\subfloat[(a)]{\includegraphics[width=0.25\textwidth]{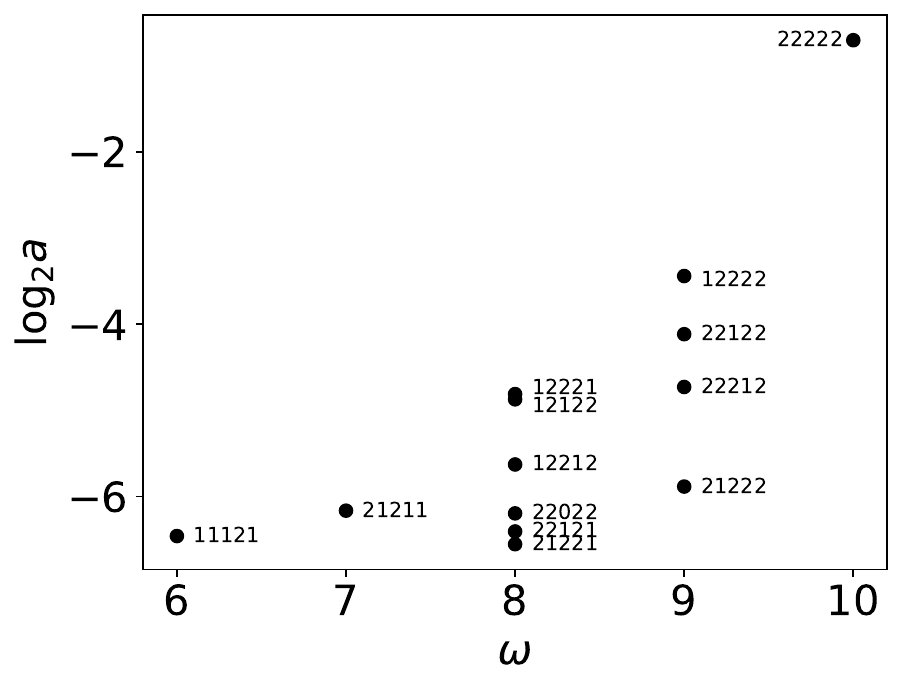}}
\subfloat[(b)]{\includegraphics[width=0.25\textwidth]{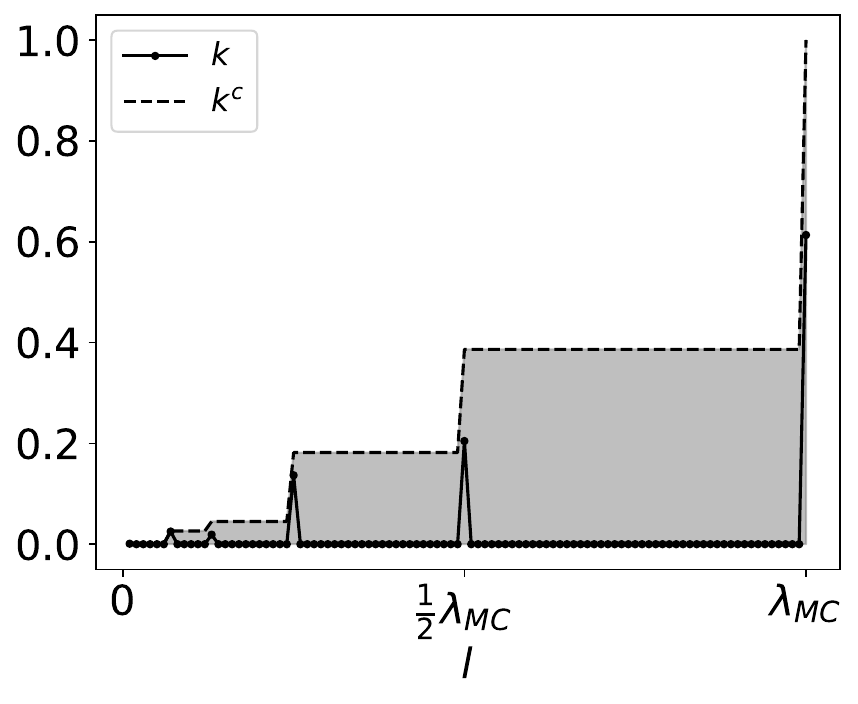}}
\subfloat[(c)]{\includegraphics[width=0.2\textwidth]{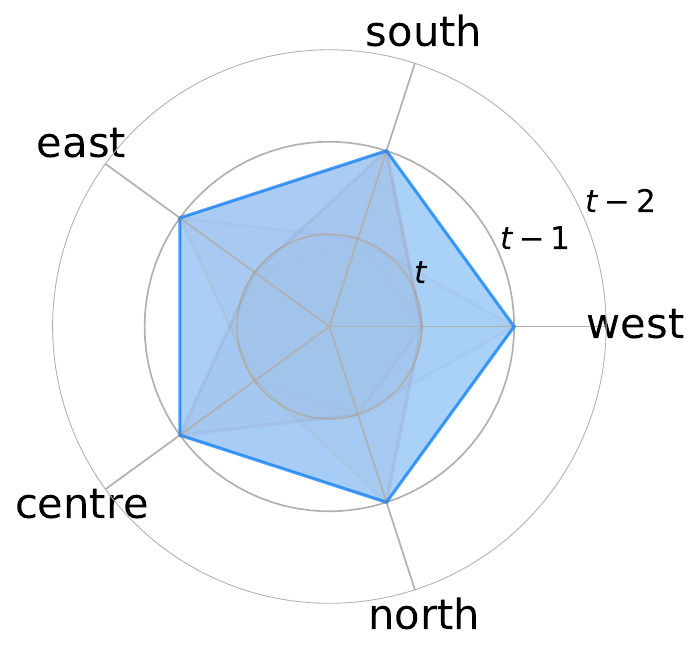}}\\[-0.25cm]
\caption{We report the results of our analysis of Traffic jams. (a) Selection of dominant interactions with $a > 0.01$ whose coefficients are plotted on a logarithmic scale and are labelled with the corresponding interaction type $\mathbf{q}$. (b) Weights and reducibility. (c) We represent most relevant interactions with their $\mathbf{q}$ in a polar plot where the radial and angular coordinates account for $q_n$ and $n$ respectively. The opacity of each interaction corresponds to its interaction coefficient.}  
\label{sup_fig:jams}
\end{figure}

Results in SM Fig. \ref{sup_fig:jams} show how the system is mainly dominated by $\interaction$ with type $q_n=2$.
A closer look at the interaction's transition matrix reveals the most important trajectories at the phase space: 
We have $\s(t-1)=(0,0,0,0,0) \rightarrow \s(t)=(0,0,0,0,0)$, $\s(t-1)=(1,1,1,1,0) \rightarrow \s(t)=(1,1,1,1,0)$ and $\s(t-1)=(1,1,1,1,1) \rightarrow \s(t)=(1,1,1,1,1)$ amongst the processes with a higher probability to occur, where the three of them express a situation in which the system remains in the same state.
Next on the list we have $\s(t-1)=(1,1,1,1,0) \rightarrow \s(t)=(1,1,1,1,1)$, $\s(t-1)=(1,1,1,1,1) \rightarrow \s(t)=(1,1,1,1,0)$ and $\s(t-1)=(1,1,1,1,0) \rightarrow \s(t)=(1,1,0,1,0)$. The first two account respectively for the propagation of a traffic jam to the north of the city when the rest of locations are also collapsed and the opposite process. The latter expresses a traffic dispersion in the east when all but the north are collapsed. Finally, despite $a$ of $q_n=2$ being the largest coefficient by far, the rest are not negligible which yields a reducibility of $r=0.26$.\\ 

\subsection{The Well-Tempered Clavier}
\begin{figure}[htb]
\captionsetup[subfigure]{labelformat=empty}
\centering
\subfloat[(a)]{\includegraphics[width=0.25\textwidth]{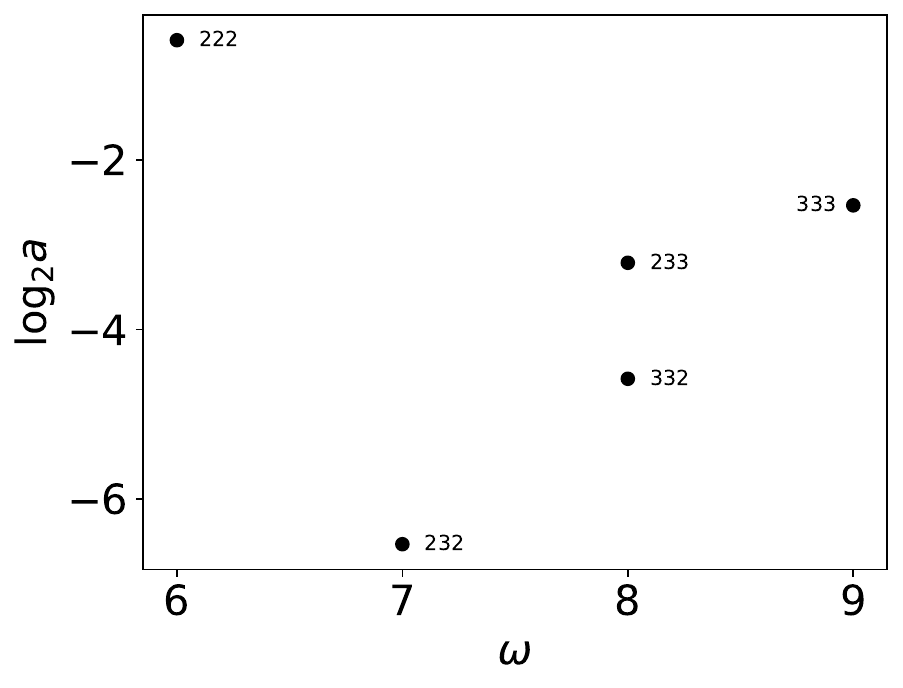}}
\subfloat[(b)]{\includegraphics[width=0.25\textwidth]{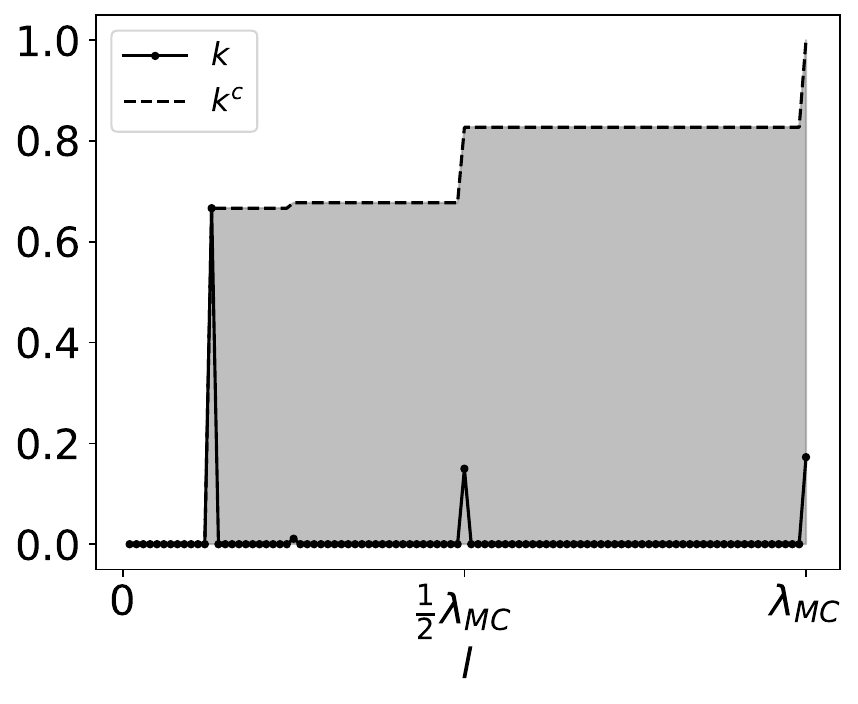}}
\subfloat[(c)]{\includegraphics[width=0.2\textwidth]{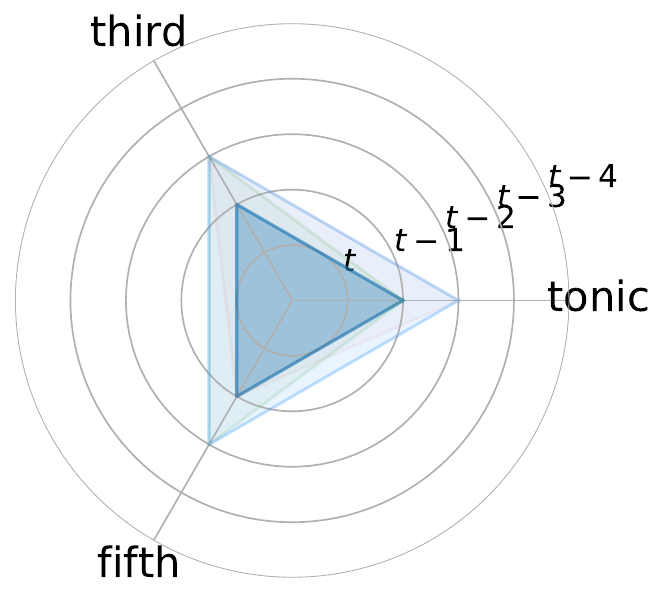}}\\[-0.5cm]
\caption{We report the results of our analysis of Musical compositions. (a) Selection of dominant interactions with $a > 0.01$ whose coefficients are plotted on a logarithmic scale and are labelled with the corresponding interaction type $\mathbf{q}$. (b) Weights and reducibility. (c) We represent most relevant interactions with their $\mathbf{q}$ in a polar plot where the radial and angular coordinates account for $q_n$ and $n$ respectively. The opacity of each interaction corresponds to its interaction coefficient.}  
\label{sup_fig:music}
\end{figure}

We have analysed a compilation of the 18 major pieces of the Well-Tempered Clavier by J.S. Bach. The possible node states are the sound or silence, $|X_n|=2$, of the tonic, third and fifth notes of the scale, $N=3$. For a maximum value of $o$ given by $\order^+=6$ the AIC establishes a value of $\order=3$ for this dataset composed by $L=1759947$ events, which yields $\eta \sim 10^{3.5}$ and an associated $\sigma(\mathbf{a},\hat{\mathbf{a}}) \sim 0.1$ if we extrapolate the numerical results from the experiments with synthetic data. See the outcome of the study at SM Fig. \ref{sup_fig:music}.\\

Interaction coefficients at SM Fig. \ref{sup_fig:music} shed light on the dominant character of interaction type $\q=(2,2,2)$.
The interaction's transition matrix is very close to the identity matrix of dimension $|X|$, which represents a simple static process of copying the previous state of the system.
We argue that the prevalence of the interaction type with $\q=(2,2,2)$ and the overall reducible character of the system ($r=0.67$) are a direct consequence of the time resolution of the dataset $\mathbf{D}$. We have obtained $\mathbf{D}$ by translating midi files, for which two consecutive time steps are separated only a tiny fraction of the duration of average notes. Thus, repetition is the standard behaviour expected in the system.

\end{document}